\UseRawInputEncoding
\documentclass[superscriptaddress, twocolumn, amsmath, amssymb, aps,pra, notitlepage,longbibliography]{revtex4-1} %preview twocolumn
\usepackage{graphicx,graphics,epsfig,subfigure,times,bm,bbm,amssymb,amsmath,amsfonts,amsthm,mathrsfs,MnSymbol,physics}
\usepackage[matrix,frame,arrow]{xypic}
\usepackage[normalem]{ulem}
\usepackage{slashed}
\usepackage{dcolumn}
\usepackage{color}
\usepackage[usenames,dvipsnames,svgnames,table]{xcolor}
\usepackage[english]{babel}
\usepackage{verbatim}
\definecolor{darkblue}{rgb}{0.0,0.0,0.3}
\usepackage[colorlinks=true,
            linkcolor=red,
            urlcolor= darkblue,
            citecolor=blue]{hyperref}

\newcommand{\bea}{\begin{eqnarray}}
\newcommand{\eea}{\end{eqnarray}}

\begin{document}
\title{Direct temperature readout in nonequilibrium quantum thermometry}
%\title{Direct temperature inference from nonequilibrium quantum thermometry: Error estimation and dephasing effect}
%\title{Nonequilibrium quantum thermometry for direct temperature inference}

\author{Yan Xie}
\affiliation{Department of Physics, Institute for Quantum Science and Technology, Shanghai Key Laboratory of High Temperature Superconductors, International Center of Quantum and Molecular Structures, Shanghai University, Shanghai, 200444, China}
\author{Junjie Liu}
\email{jj\_liu@shu.edu.cn}
\affiliation{Department of Physics, Institute for Quantum Science and Technology, Shanghai Key Laboratory of High Temperature Superconductors, International Center of Quantum and Molecular Structures, Shanghai University, Shanghai, 200444, China}

\begin{abstract}
Quantum thermometry aims to measure temperature in nanoscale quantum systems, paralleling classical thermometry. However, temperature is not a quantum observable, and most theoretical studies have therefore concentrated on analyzing fundamental precision limits set by the quantum Fisher information through the quantum Cram\'er-Rao bound. In contrast, whether a direct temperature readout can be achieved in quantum thermometry remains largely unexplored, particularly under the nonequilibrium conditions prevalent in real-world applications. To address this, we develop a direct temperature readout scheme based on a thermodynamic inference strategy. The scheme integrates two conceptual developments: (i) By applying the maximum entropy principle with the thermometer's mean energy as a constraint, we assign a reference temperature to the nonequilibrium thermometer. We demonstrate that this reference temperature outperforms a commonly used effective temperature defined through equilibrium analogy. (ii) We obtain positive semi-definite error functions that lower-bound the deviation of the reference temperature from the true temperature and vanish upon thermalization with the sample. Combining the reference temperature with these error functions, we introduce a notion of corrected dynamical temperature which furnishes a postprocessed temperature readout under nonequilibrium conditions. This corrected dynamical temperature can be evaluated adaptively without prior knowledge of the actual temperature. We validate the corrected dynamical temperature in a qubit-based thermometer under a range of nonequilibrium initial states, confirming its capability to estimate the true temperature. Importantly, we find that increasing quantum coherence can enhance the precision of this readout. Our findings complement existing research on quantum thermometry and help bridge the gap between prevailing theoretical analysis on precision limit and the practical need of direct temperature readout.
\end{abstract}

\maketitle
\date{\today}

\section{Introduction}
Quantum thermometry~\cite{Mehboudi.19.JPA} serves a dual and indispensable role in advancing quantum science. Fundamentally, it provides the essential tool to quantify temperature--a central thermodynamic parameter--thereby enabling the investigation of thermodynamic processes and thermal machines in genuinely quantum regimes~\cite{Hartmann.04.PRL,Maruyama.09.RMP,Parrondo.15.NP,Goold.16.JPA}. Practically, precise temperature measurement is critical for many quantum technologies, particularly in quantum simulation~\cite{Georgescu.14.RMP,Bloch.08.RMP} and quantum information processing~\cite{Wendin.17.RPP}, where physical systems are typically initialized and maintained at cryogenic temperatures.
Real-time and accurate temperature monitoring during cooling or quantum control processes is essential to achieve and verify the requisite operational conditions for quantum devices~\cite{Olf.15.NP,Hohmann.16.PRA,Lous.17.PRA,Mecklenburg.15.S}.

A standard approach to quantum thermometry is probe-based thermometry~\cite{Brunelli.11.PRA,Brunelli.12.PRA,Mitchison.20.PRL,Mehboudi.19.PRL,Mehboudi.19.JPA,Rubio.21.PRL,Jevtic.15.PRA,Correa.15.PRL,Ullah.25.PRA,Aiache.24.PRA,Albarelli.23.PRA,Zhang.22.NPJQI,Razavian.19.EPJP,Candeloro.21.PRA,Ullah.23.PRR,Cruzeiro.24.E,Hovhannisyan.21.PRXQ,Segal.24.PRA,Seah.19.PRL,Hohmann.16.PRA,Lous.17.PRA,Jorgensen.20.PRR,Aiache.25.PRA}, where the use of miniaturized systems as measurement probes can minimize perturbative effects on the measured sample. Within this framework, temperature is estimated indirectly from the knowledge of the state of a given probe that undergoes an evolution process after coupling to the sample. The typical metrological protocol consists of three stages: the probe is prepared in a specific initial state; it undergoes an evolution that encodes information about the sample’s temperature; finally, a suitable measurement is performed on the probe to extract that information. In the ideal scenario where the thermometry fully thermalizes with the sample, temperature can be estimated with optimal accuracy from its mean energy~\cite{Olf.15.NP,Correa.15.PRL,Hohmann.16.PRA,Lous.17.PRA,Mehboudi.19.PRL,Hovhannisyan.21.PRXQ,Hovhannisyan.18.PRB,Jorgensen.20.PRR}. However, achieving thermal equilibrium can be experimentally challenging at low temperatures, where thermalization times are relatively long compared with typical measurement durations. Consequently, non-equilibrium thermometry where measurements are performed on out-of-equilibrium probes has emerged as a vital alternative~\cite{Jevtic.15.PRA,Kiilerich.18.PRA,Segal.24.PRA,Cavina.18.PRA,Sekatski.22.Q,Brenes.23.PRA,Ullah.25.PRA,Aiache.25.PRA}. 

Despite significant progress, existing theoretical analyses of quantum thermometry have predominantly focused on the quantum Fisher information (QFI)~\cite{LiuJ.19.JPA}, which determines the fundamental limit of the mean squared error through the celebrated quantum Cram\'er-Rao bound (QCRB)~\cite{Braunstein.94.PRL}. While QFI analysis provides valuable insights for optimizing metrological performance, its practical relevance can be potentially hindered under realistic experimental conditions--such as finite data sets and suboptimal measurements--where the saturation of the QCRB is often unattainable. Moreover, existing QFI-based analyses primarily aim to minimize the attainable mean squared error by maximizing the QFI, rather than addressing the issue of reconstructing temperature value from measurement data. In contrast, practical applications of thermometry generally demand a direct readout of temperature itself~\cite{Brites.12.Na,Mecklenburg.15.S,Olf.15.NP,Hohmann.16.PRA,Lous.17.PRA,YeL.16.PRB,Zhang.19.PR,Goblot.26.PRL}, rather than a bound on the second-order mean squared error. This disparity underscores a notable gap between the prevailing theoretical emphasis on precision limits and the practical demands of temperature readout.

Yet, temperature readout in quantum nonequilibrium thermometry is not as straightforward as in its classical counterpart. First, temperature is not a directly measurable observable; it can only be inferred indirectly from measuring suitable observables. Ideally, if the functional dependence of the mean value of a chosen observable on temperature is known, one can simply construct a mapping from measurement outcomes to temperature values. However, except in exactly solvable models, such a functional dependence is often unknown {\it a priori}--the governing dynamics typically involve multiple parameters beyond temperature, let alone the complexities encountered in realistic experiments. Consequently, constructing an exact mapping between an observable-based estimator and temperature is difficult, especially when just limited data is available. Second, strictly speaking, temperature is thermodynamically well-defined only when the probe reaches thermal equilibrium with the sample. Hence, assigning a thermodynamic temperature to a nonequilibrium quantum thermometry undergoing finite-time processes remains a general challenge~\cite{Zhang.19.PR}. Addressing the gap outlined above calls for theoretical tools that can circumvent these difficulties.

In this work, we bridge this gap by introducing a direct temperature readout scheme for nonequilibrium quantum thermometer. We consider a finite-sized quantum thermometer with a known Hamiltonian that is weakly coupled to a thermal sample, such that measurement back-action on the sample is suppressed and the thermal equilibrium state of the thermometer takes a Gibbsian form. Our scheme utilizes just the mean energy of the thermometer and provides a postprocessed temperature estimate at any given time that is guaranteed to converge to the true temperature upon thermalization. Our scheme addresses the aforementioned difficulties by integrating two conceptual developments: (i) Recognizing that temperature is not a direct observable and that experimental data are often limited, we develop a thermodynamic inference strategy suited for finite data sets. This strategy assigns a time-dependent reference Gibbsian state, and hence a reference temperature, to the nonequilibrium thermometer by applying the maximum entropy principle~\cite{Jaynes.57.PR} with the thermometer's instantaneous mean energy as a dynamical constraint. We demonstrate that this reference temperature outperforms a commonly employed effective temperature defined through equilibrium analogy~\cite{Zhang.19.PR,Alipour.21.A,Bartosik.23.PRL,Chatterjee.23.PRL,Burke.23.PRE,Sorkin.24.PRL} in Markovian thermal relaxation processes, thereby endowing it with clear thermodynamic relevance as a physically meaningful effective temperature. (ii) We further introduce positive semi-definite error functions that lower-bound the deviation of the reference temperature from the true temperature.
%analogous to how the QCRB lower-bounds the second-order mean squared error. 
These error functions vanish when the thermometer thermalizes with the sample. The final postprocessed temperature readout, dubbed corrected dynamical temperature, is then obtained by correcting the reference temperature using the associated error.

We emphasize that the proposed readout scheme is experimentally feasible for a finite-sized thermometer, as it requires just the knowledge of the thermometer's state to evaluate mean energy. This can be accessed via quantum state tomography which has been implemented in systems of up to tens of qubits~\cite{Lanyon.17.NP,Yoneda.21.NC,Zhong.21.N}. Moreover, we develop an iterative procedure for evaluating the corrected dynamical temperature when prior knowledge of the actual temperature is unavailable, marking a departure from local thermometry. We validate the scheme using a qubit-based quantum thermometer as an illustrative example, demonstrating that the corrected dynamical temperature delivers a reliable estimation of the actual temperature. We show that the accuracy of this readout can be enhanced through an initial-state engineering by tuning the population and coherence of the probe's initial state. We remark that our scheme complements existing methodologies. If a priori knowledge of the temperature is available, one can first determine an optimal thermometry setup via QFI analysis and subsequently employ our scheme for direct readout. In cases where no such knowledge exists, our scheme can be implemented iteratively, similar in spirit to global thermometry.

The structure of this paper is as follows. In Section~\ref{sec:1}, we outline the general temperature readout scheme for a nonequilibrium quantum thermometer. This includes introducing a reference Gibbs state and the corresponding reference temperature, defining suitable error functions, and formulating the corrected dynamical temperature that serves as the final postprocessed temperature readout. Section~\ref{sec:2} illustrates the scheme by focusing on a qubit-based quantum thermometer. We first analyze its QFI to benchmark its performance as a thermometer, in line with existing studies. We then examine the behavior of the reference temperature and error functions, evaluate the postprocessed temperature readout for various nonequilibrium initial states, and identify strategies for enhancing readout accuracy. Finally, in Section~\ref{sec:3}, we summarize the study with concluding remarks. Derivation details are provided in the appendices.

%==================================================================
\section{Direct temperature readout scheme}\label{sec:1}
In this section, we present the essential components of a direct temperature readout scheme for probe-based quantum thermometry. First, we develop a thermodynamic inference method based on the maximum entropy principle~\cite{Jaynes.57.PR}, which assigns a reference Gibbsian state--and thus a reference temperature--to the nonequilibrium quantum thermometer. We show that this reference temperature carries thermodynamic relevance and can be interpreted as an effective temperature. We then derive general positive semi-definite lower bounds on the deviation of the reference temperature from the actual temperature; these bounds constitute the error functions necessary to assess the performance of the thermodynamic inference. Finally, we introduce a corrected dynamical temperature that serves as the final postprocessed temperature readout for the nonequilibrium thermometer. To maintain generality, no specific assumptions are made at this stage regarding the detailed properties of the thermometer, including its Hamiltonian $H_p$ and evolution dynamics. We set $k_B=1$ and $\hbar=1$ hereafter.

%==================================================================
\subsection{Thermodynamic inference and reference temperature}\label{sec:inference}
To establish our framework, we assume that the measured sample remains in a thermal equilibrium state $\rho_T=e^{-H_p/T}/\mathrm{Tr}[e^{-H_p/T}]$ characterized by a well-defined thermodynamic temperature $T$, which is the parameter to be estimated by using a quantum thermometer. For convenience, we denote the inverse temperature as $\beta=T^{-1}$. If this assumption is not met, the task of quantum thermometry becomes ill-posed. In contrast, the quantum thermometer coupled to the sample may reside in a nonequilibrium state during finite-time evolution. The objective of nonequilibrium quantum thermometry is thus to estimate the sample temperature from the nonequilibrium dynamics of the thermometer.

To this end, one can measure observables of the nonequilibrium thermometer. Since the functional dependence of the mean values of observables on temperature is generally unknown, the problem necessitates inferring temperature from measurements, a process inherently subject to statistical error. To establish an inference strategy that is both operationally robust and thermodynamically consistent, we need to address two basic questions: (i) Which observable should be selected? (ii) How can we ensure that the inferred parameter is physically identifiable as a temperature? The maximum entropy principle~\cite{Jaynes.57.PR} provides a key conceptual basis for this: When the system's mean energy is the only available observation, the least-biased inference of the system's state is a Gibbsian state. Consequently, by choosing the Hamiltonian as the relevant observable, the maximum entropy principle guarantees an inferred state characterized by a parameter with the dimensions of temperature. Notably, this inference naturally reduces to the exact result under conditions of thermal equilibrium.

Building on this conceptual fit, we apply the maximum entropy principle, originally developed for thermal equilibrium systems~\cite{Jaynes.57.PR}, to nonequilibrium settings and propose a temperature inference strategy tailored to nonequilibrium quantum thermometry. Specifically, following the intuition of the maximum entropy principle~\cite{Jaynes.57.PR}, we assign a time-dependent Gibbsian state $\rho_r(t)=e^{-\beta_r(t)H_p}/Z_r(t)$ with $Z_r(t)=\mathrm{Tr}[e^{-\beta_r(t)H_p}]$ the partition function to the nonequilibrium thermometer. This Gibbsian state is fixed by the following dynamical constraint imposed by the instantaneous mean energy of the nonequilibrium thermometer~\cite{MaT.19.PRA,Strasberg.21.PRXQ,LiuJ.24.PRR}
\begin{equation}\label{eq:equal_e}
    E_p(t)~\equiv~\mathrm{Tr}[H_p\rho_p(t)]~=~\mathrm{Tr}[H_p\rho_r(t)].
\end{equation}
Here, $\rho_p(t)$ is the actual nonequilibrium state of the quantum thermometer. $E_p(t)$ uniquely determines the value of $\beta_r(t)$. We stress that $\rho_r(t)$ needs not possess a meaningful thermodynamic interpretation on its own: rather, its foundation rests in experimental observations on thermometer's energetics. If the thermometer evolves quasi-statically along an instantaneous equilibrium path, $\rho_r(t)$ coincides with the actual state $\rho_p(t)$ of the thermometer. Beyond this special case, $\rho_r(t)$ generally deviates from $\rho_p(t)$ and can only be treated as a {\it reference state that is least biased while remaining compatible with the energy constraint. In this sense,} we identify $\beta_r(t)$ as the inverse reference temperature inferred from the energetics of the quantum thermometer. This maximum-entropy inference approach aligns quantum thermometry more closely with its classical counterpart and offers a concrete route towards a practical temperature readout.
Notably, this energy-based strategy for obtaining a reference temperature has recently been demonstrated experimentally~\cite{Aimet.25.NP}.

To further assess the thermodynamic relevance of the reference temperature, we can consider an illustrating scenario in which nonequilibrium thermometers undergo Markovian thermal relaxation processes towards the final thermalization with the sample. This is possible as we consider weak probe-sample couplings. For this setting, we analytically prove that the deviation of the inverse reference temperature from the true inverse temperature satisfies the inequality
\begin{equation}\label{eq:beta_dif}
|\beta-\beta_r(t)|~\le~|\beta-\beta_e(t)|.
\end{equation}
For clarity, we relegate derivation details to Appendix~\ref{a:1}. Here, $\beta_e(t)\equiv\left[\partial E_p(t)/\partial S(t)\right]^{-1}$ with $S(t)=-\mathrm{Tr}[\rho_p(t)\ln\rho_p(t)]$ the actual von Neumann entropy of the probe is a widely used effective temperature definition for nonequilibrium systems~\cite{Zhang.19.PR,Alipour.21.A,Bartosik.23.PRL,Chatterjee.23.PRL,Burke.23.PRE,Sorkin.24.PRL}--a direct generalization of the equilibrium thermodynamic definition. The significance of Eq. (\ref{eq:beta_dif}) is that the inverse reference temperature $\beta_r(t)$ obtained from a nonequilibrium thermodynamic inference is a more accurate estimate of the true inverse temperature at finite times than $\beta_e(t)$. This result underscores the utility of the reference temperature as a well-founded effective temperature in nonequilibrium settings.

Thus, by utilizing the mean energy of the nonequilibrium thermometer as the available observation, we can infer a reference temperature that has a thermodynamic significance. However, this reference temperature alone does not yet furnish the final temperature readout of nonequilibrium thermometers. Its value is essentially fixed by the underlying energetic dynamics of the probe, leaving little room to improve readout performance other than by blindly tuning those energetic dynamics without guiding principles. Consequently, a lower bound on the temperature deviation $|\beta-\beta_r(t)|$--which directly quantifies the error of the thermodynamic inference--would be more informative than the upper bound provided in Eq.~\eqref{eq:beta_dif}, especially given that the latter is restricted to Markovian relaxation processes. Such a lower bound enables us to assess the attainable accuracy of the thermometer in estimating the temperature value, an issue we address in the following subsection.

%==================================================================
\subsection{Lower bounds on the temperature deviation}\label{sec:bounds}
To endow the direct temperature readout strategy with operational significance, we now seek lower bounds on the deviation of the reference temperature from the true temperature. We recall that the reference temperature is derived from a thermodynamic inference model that uses only the mean energy of the thermometer. Thus, lower-bounding the temperature deviation can be approached by taking into account either (i) the accuracy of this inference model in reconstructing the true probe state, or (ii) the deviation of the instantaneous mean energy from its equilibrium value. In both cases, as the thermometer approaches thermal equilibrium with the sample, these deviations vanish, and the reference temperature converges to the true sample temperature. This observation suggests that meaningful lower bounds on the temperature deviation can be formulated in terms of either model accuracy or energy-based discrepancy. For simplicity, time-dependence is suppressed in this subsection.

We note that the difference in von Neumann entropy $S_r-S\ge 0$ ($S_r=-\mathrm{Tr}[\rho_r\ln\rho_r]$) naturally reflects the accuracy of the thermodynamic reference model building upon the maximum entropy principle. Since $S_r-S=D(\rho_p||\rho_r)$~\cite{Xiao.25.A}, with $D(\rho_p||\rho_r)=\mathrm{Tr}[\rho_p(\ln\rho_p-\ln\rho_r)]$ being the quantum relative entropy, we expect that the quantum relative entropy $D(\rho_p||\rho_r)$ provides a natural candidate for bounding the temperature deviation. To proceed, we utilize a generalized definition of the nonequilibrium free energy $\mathcal{F}= F_r +T_r D(\rho_p||\rho_r)$~\cite{Liu.23.PRAa}, where $F_r=-T_r\ln Z_r$. This expression allows us to express the temperature deviation in terms of entropic terms $(T_r - T) S_r = T_r D(\rho_p || \rho_r) + (T_r S - T S_r)$ (see details in Appendix~\ref{a:2}). From this relation, we can get a lower bound on the absolute temperature deviation (Appendix~\ref{a:2})
\begin{equation}\label{eq3}
|T_r - T| ~\ge~ \left| \frac{T_r D(\rho_p||\rho_r)}{S_r}-\left|T_r\frac{S}{S_r}-T\right| \right|.
\end{equation}
Clearly, this lower bound is positive semi-definite and vanishes only when the system reaches thermal equilibrium at which we have $\rho_p=\rho_r$ and $\beta_r=\beta$ (Recalled that we consider weak probe-sample couplings).

Alternatively, we can relate temperature deviation to the deviation of the instantaneous energy from its equilibrium value. Introducing an interpolating inverse temperature $\beta_s\equiv\beta+s(\beta_r-\beta)$ with $s\in[0,1]$, we can define a corresponding Gibbsian state $\rho_g^s\equiv e^{-\beta_sH_p}/Z_s$ with $Z_s=\mathrm{Tr}[e^{-\beta_sH_p}]$. This construction yields
\bea\label{eq:contrast_def}
E_T-E_p &=& -\mathrm{Tr}\left[\int_0^1\frac{d}{ds}\left(H_p\rho_g^s\right)ds\right].
\eea
Here, $E_T=\mathrm{Tr}[H_p\rho_T]$ is the internal energy of the quantum thermometer in thermal equilibrium. From the above relation, we can obtain a lower bound on $|\beta_r-\beta|$ expressed solely in terms of the thermometer's energetics~(see details in Appendix~\ref{a:3})
\begin{equation}\label{eq:beta_bound}
    |\beta_r-\beta| ~\ge~ \frac{|E_T-E_p|}{(||H_p||_{\infty})^2}.
\end{equation}
Here, $||H_p||_{\infty}$ denotes the operator norm of the probe Hamiltonian, which for a Hermitian operator equals its largest absolute eigenvalue. Similar to Eq. (\ref{eq3}), we see that this lower bound is also positive semi-definite and vanishes upon thermalization with the sample where we have $E_T=E_p$.

We emphasize that inequalities Eqs. (\ref{eq3}) and (\ref{eq:beta_bound}) impose general constraints on the temperature deviation, independent of the specific details of the quantum thermometer and its nonequilibrium dynamics. They establish well‑defined ultimate limits for the temperature-readout error via a thermodynamic inference strategy. For later convenience, we refer to these lower bounds explicitly as error functions
\bea
\mathcal{E}_1 &\equiv& \left| \frac{T_r D(\rho_p||\rho_r)}{S_r}-\left|T_r\frac{S}{S_r}-T\right| \right|, \label{eq:epson1}\\
\mathcal{E}_2 &\equiv& \frac{|E_T-E_p|}{(||H_p||_{\infty})^2}.\label{eq:epson2}
\eea
Both are positive semi-definite, as analyzed before.

Before proceeding, several remarks concerning these error functions are in order: (i) $\mathcal{E}_1$ and $\mathcal{E}_2$ apply to the deviations of temperature and inverse temperature, respectively. They are complementary rather than equivalent. (ii) Evaluating $\mathcal{E}_1$ and $\mathcal{E}_2$ requires knowledge of the probe’s Hamiltonian and state. While the former is usually assumed to be known, the latter can be accessed by resorting to quantum state tomography whose experimental overhead for a finite-sized quantum thermometer is compatible with current experimental capabilities~\cite{Lanyon.17.NP,Yoneda.21.NC,Zhong.21.N}. (iii) We note $\mathcal{E}_1$ depends on thermometer's quantum coherence defined in the energy basis of $H_p$, as it involves the full state and entropy. By contrast, $\mathcal{E}_2$ involves only the thermometer's energy which is determined solely by the populations of the thermometer's state in the energy basis. For Markovian dynamics governed by a quantum Lindblad master equation--where populations and coherences evolve independently~\cite{Breuer.02.NULL}--$\mathcal{E}_2$ is therefore likely to be insensitive to coherence. (iv) These error functions do not vanish when the thermometer is out of thermal equilibrium, irrespective of the number of measurements. In our inference scheme, only the mean energy of the thermometer is relevant; once this quantity is determined, it cannot be altered by increasing the number of measurements. This property is in direct contrast to the QCRB, where the mean squared error can vanish in the asymptotic limit of infinitely many measurements.

%==================================================================
\subsection{Final postprocessed temperature readout}
Building on the concepts of a reference temperature $T_r(t)$ (or its inverse $\beta_r(t)$) and its associated error functions $\mathcal{E}_{1}(t)$ (or $\mathcal{E}_{2}(t)$), we now formulate a practical scheme for direct temperature readout in nonequilibrium quantum thermometry. This scheme integrates the thermodynamic inference introduced in Sec.~\ref{sec:inference} with the error bounds derived in Sec.~\ref{sec:bounds} to produce a postprocessed, time-dependent temperature estimate--dubbed the corrected dynamical temperature--that is both experimentally accessible and theoretically grounded. Below, we present its construction and explain its practical implementation.

We note that the error functions $\mathcal{E}_1(t)$ and $\mathcal{E}_2(t)$ establish exclusion deviations from the true temperature. For instance, the inequality Eq. (\ref{eq3}) implies that $T_r(t)$ must differ from $T$ by at least $\mathcal{E}_1(t)$ at finite times, with the direction of the deviation $T_r(t)-T\le -\mathcal{E}_1(t)$ or $T_r(t)-T\ge \mathcal{E}_1(t)$ determined by whether the probe undergoes heating or cooling upon coupling to the thermal sample at $t=0$, respectively. A similar relation follows from Eq. (\ref{eq:beta_bound}) for the inverse temperature. This structure motivates the introduction of the corrected dynamical temperature, defined as the reference temperature shifted by the corresponding error bound
\bea
    T_{\rm{corr}}(t) &\equiv& T_r(t)+\chi_1\mathcal{E}_{1}(t),\label{eq:corr_T}\\
    \beta_{\rm{corr}}(t) &\equiv& \beta_r(t)+\chi_2\mathcal{E}_{2}(t).\label{eq:corr_beta}
\eea
Here, $\chi_{1,2}\in\{+1,-1\}$ are coefficients whose values are fixed by the thermodynamics of the thermal relaxation process as we will explain below. We remark that $\beta_{\rm{corr}}(t)$ is not the inverse of $T_{\rm{corr}}(t)$; the two quantities are independently defined and provide complementary readouts. In practical implementation, the sign of $\chi_{1,2}$ is determined by the initial energy of the thermometer relative to its equilibrium value $E_T$, which dictates the direction of energy flow and thus whether $T_r(t)$ approaches $T$ from above or below:
\begin{itemize}
    \item [(i)] In the cooling regime with $E_{p}(0)>E_{T}$, the thermometer releases energy into the thermal sample, so $T_r(t)>T$ during the relaxation. The error bound then forces the true temperature to lie below the reference estimate, leading to natural choices of $\chi_1=-1$ in Eq.~(\ref{eq:corr_T}) and $\chi_2=1$ in Eq.~(\ref{eq:corr_beta}),
    \bea
    T_{\rm{corr}}(t) &\equiv& T_r(t)-\mathcal{E}_{1}(t),\\
    \beta_{\rm{corr}}(t) &\equiv& \beta_r(t)+\mathcal{E}_{2}(t).
    \eea
    \item [(ii)] In the heating regime with $E_{p}(0)<E_{T}$, the thermometer absorbs energy from the thermal sample, giving $T_r(t)<T$. In this case, the true temperature must lie above the reference window, leading to the opposite sign assignment
    \bea
    T_{\rm{corr}}(t) &\equiv& T_r(t)+\mathcal{E}_{1}(t),\\
    \beta_{\rm{corr}}(t) &\equiv& \beta_r(t)-\mathcal{E}_{2}(t).
    \eea
\end{itemize}

Physically, the corrected dynamical temperature--whether expressed as $T_{\rm{corr}}(t)$ or $\beta_{\rm{corr}}(t)$--represents a thermodynamically consistent postprocessing of the raw reference temperature, providing the final postprocessed temperature readout in nonequilibrium quantum thermometry. It explicitly incorporates the direction of thermalization through the sign of the error shift, ensuring that the readout converges monotonically to the true temperature as the thermometer equilibrates. This construction bridges the gap between the instantaneous reference temperature (which alone lacks an intrinsic accuracy measure) and an operationally meaningful temperature estimate endowed with a built-in error bound.

We note that both error functions, as evident from their expressions, depend on the actual temperature, similar to the QFI in the QCRB. This appears to confine our scheme to the framework of local thermometry, which requires prior knowledge of the actual temperature. However, when such prior knowledge is unavailable, our scheme can still be implemented iteratively by treating $T$ in the error functions $\mathcal{E}_{1,2}(t)$ as an iterative parameter. To be precise, we detail the first step of the iteration procedure for evaluating $T_{\rm{corr}}(t)$ as an example: First, one starts with an initial guess $T_{\rm{init}}$ at $t=0$, which serves as $T$ in the expression of the error function $\mathcal{E}_1(0)$. Second, one evaluates the initial reference temperature $T_r(0)$ using the knowledge of the initial state. Third, one computes $\mathcal{E}_1(0)$ from $T_{\rm{init}}$ and $T_r(0)$, and subsequently obtains $T_{\rm{corr}}(0)$. To complete the next step, one simply uses $T_{\rm{corr}}(0)$ as the updated guess of the actual temperature and repeats the first step to obtain $T_{\rm{corr}}(\Delta t)$, where $\Delta t$ is the time step. Repeating these steps yields an iterative evaluation of $T_{\rm{corr}}(t)$.

In the following section, we demonstrate this direct temperature readout scheme using a qubit-based thermometer and show how initial-state engineering, especially through the tuning of initial quantum coherence and population, can improve the accuracy of the postprocessed temperature readout.

%==================================================================
\section{Example: Qubit-based quantum thermometer}\label{sec:2}
In this section, we validate the proposed direct temperature readout framework by applying it to a paradigmatic qubit-based quantum thermometer~\cite{Jevtic.15.PRA,Correa.15.PRL,Mehboudi.19.JPA,Ullah.25.PRA} whose fabrication is well within the current experimental capacities~\cite{Kuffer.25.PRXQ}. Our analysis is structured to systematically demonstrate the validity of the scheme and to identify the physical resources that improve its precision. First, to establish a benchmark within the conventional QCRB framework, we analyze the behavior of the QFI $\mathcal{F}_T$ of the system. We then proceed to implement our readout scheme and investigate whether and how an initial-state engineering can enhance the resulting readout precision. Except in Sec.~\ref{sec:d3}, we assume that prior knowledge of the actual temperature is available.

%==================================================================
\subsection{Model}\label{sec:model}
To estimate the temperature $T$ of the thermal sample using a qubit probe, we couple the qubit to the sample and model its dissipative evolution via the following quantum Lindblad master equation~\cite{Breuer.07.NULL}
\begin{equation}\label{eq:master}
    \partial_t\rho_p(t)~=~-i[H_p,\rho_p(t)]+\sum_{\mu}\gamma_{\mu}\mathcal{D}[J_{\mu}] \rho_p(t).
\end{equation}
Here, $\partial_t=\partial/\partial t$ denotes the time derivative, $H_p=\omega\sigma_z /2$ is the qubit Hamiltonian with energy gap $\omega$ and Pauli-Z matrix $\sigma_z$, and the dissipation experienced by the qubit probe is captured by the Lindblad dissipator with $\mathcal{D}[J_{\mu}] \rho_p(t) = J_{\mu} \rho_p(t) J_{\mu}^\dagger-\{J_{\mu}^\dagger J_{\mu}, \rho_p(t)\}/2$; where $J_{\mu}$ denotes a Lindblad jump operator of dissipation channel $\mu$ with the corresponding damping strength $\gamma_{\mu}$, and $\{A,B\}=AB+BA$. 

To make a probe-based thermometry practical, it is important to comprehensively account for dissipation effects arising from the coupling of the probe to the sample, as well as from any parasitic environments. Here, we consider the simultaneous action of three realistic dissipation channels: (i) $J_+=\sigma_+$ and $J_-=\sigma_-$ describe excitation and de-excitation processes induced by the thermal sample with $\sigma_{\pm}$ the spin ladder operators. Their damping rates are $\gamma_+ = \gamma N$ and $\gamma_-=\gamma(N+1)$, where $N=1/(e^{\beta\omega} -1)$ is the Bose-Einstein distribution. (ii) $J_z=\sigma_{z}$ models a pure dephasing effect with dephasing rate $\gamma_0$. We remark that the thermal state $\rho_T=e^{-\beta H_p}/\mathrm{Tr}[e^{-\beta H_p}]$ is the unique steady state of the thermal relaxation process described by Eq. (\ref{eq:master}) in the long time limit.

Before proceeding, we highlight the distinction of our thermometric model [cf. Eq. (\ref{eq:master})] from those commonly employed in the literature. We consider coexisting effects of energy-exchanging process between the probe and the sample as well as dephasing process that is ubiquitous for qubits, whereas existing studies often treated these two processes in isolation~\cite{Ullah.25.PRA,Aiache.24.PRA,Albarelli.23.PRA,Zhang.22.NPJQI,Razavian.19.EPJP,Mitchison.20.PRL,Candeloro.21.PRA}. Moreover, we explicitly include an intrinsic dephasing channel that persists even without coupling to the sample, and we take its strength $\gamma_0$ to be temperature-independent. With the quantum Lindblad master equation Eq. (\ref{eq:master}), the time-evolving reduced density matrix of probe $\rho_p(t)$ encodes temperature information which enables us to estimate the actual temperature at finite times.

%==================================================================
\subsection{QFI characteristics}\label{sec:QFI}
To align with established literature, we first evaluate the performance of the qubit probe as a quantum thermometer within the conventional framework of quantum metrology~\cite{Braunstein.94.PRL,paris2009quantum,Helstrom.76.NULL,Holevo.11.NULL}. Within this framework, the fundamental precision limit is set by the QCRB~\cite{Braunstein.94.PRL,paris2009quantum,Helstrom.76.NULL,Holevo.11.NULL}. Analyzing this theoretical baseline allows us to rigorously quantify how quantum resources such as coherence enhance thermometric sensitivity in the presence of noise, thereby establishing clear benchmarks against which the performance of our direct readout scheme can be assessed in later sections.

To specialize the QCRB to thermometry, it sets a lower bound on the mean squared error, which reduces to the variance $ \mathrm{Var}[\mathcal{T}]$ of any unbiased temperature estimator $\mathcal{T}$ for the true temperature $T$,
\begin{equation}\label{eq:crb}
    \mathrm{Var}[\mathcal{T}] \geq \frac{1}{N \mathcal{F}_{T}}.
\end{equation}
Here, $N$ is the number of measurements, and $\mathcal{F}_{T}$ is the associated QFI about temperature $T$ defined as
\begin{equation}
    \mathcal{F}_{T}~\equiv~\text{Tr}\left[ L_{T}^2 \rho_{p} \right],
\end{equation}
%===============================================
\begin{figure}[tbh!]
 \centering
\includegraphics[width=1\columnwidth]{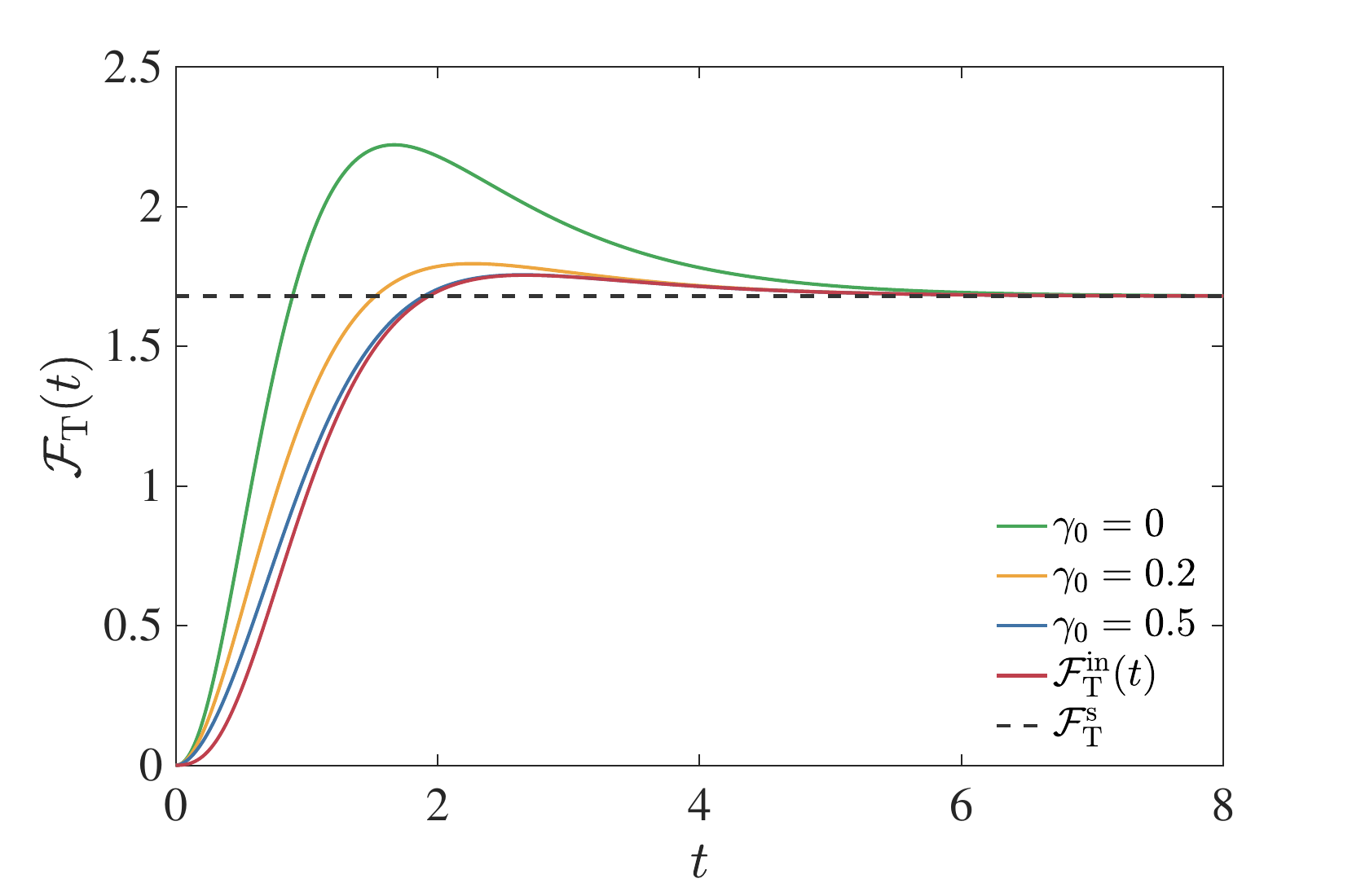} 
 \caption{Dynamics of $\mathcal{F}_T(t)$ with dephasing strength $\gamma_0=0$ (green line), $\gamma_0=0.2$ (orange line) and $\gamma_0=0.5$ (blue line) starting from a coherent initial state $\rho_p(0)=0.5\mathrm{I}+0.4\sigma_x-0.2\sigma_z$ with $\mathrm{I}$ the identity matrix. For comparison, the red curve shows $\mathcal{F}^{\rm{in}}_T(t)$ for an incoherent initial state $\rho_p(0)=0.5\mathrm{I}-0.2\sigma_z$. The black dashed line marks the value of thermal QFI given by Eq. (\ref{eq:F_T_s}) which equals the stationary value of both $\mathcal{F}_T(t)$ and $\mathcal{F}^{\rm{in}}_T(t)$ as $t\to \infty$. Other parameters are $\omega=1$, $T=0.5$ and $\gamma=1$.}
\protect\label{fig:QFI}
\end{figure}
%===============================================
where $L_{T}$ is the corresponding symmetric logarithmic derivative operator satisfying the equation $2 \partial_{T} \rho_{p} = (L_T \rho_p + \rho_p L_T)$ with $\partial_T\equiv\partial /\partial T$. The QFI $\mathcal{F}_T$ quantifies the amount of information about the temperature $T$ that can be extracted from the thermometer's actual state $\rho_p$. From a geometric perspective, the QFI also measures how sensitively the state changes under a small variation of the unknown parameter, as it is closely related to the Bures metric and the quantum Uhlmann fidelity for mixed states~\cite{Sidhu.20.AVSQ}. Notably, Eq. (\ref{eq:crb}) defines the ultimate lower limit of the mean squared error. Its saturation can be achieved by using the maximum likelihood estimator~\cite{Mehboudi.25.PRA}. In the limit of infinitely many measurement with $N\to \infty$, the minimum mean squared error can vanish, implying perfect temperature estimation. However, in realistic experiments one can usually perform only a finite number of measurements, rendering the lower bound in the QCRB finite. In this case, one cannot directly obtain a temperature value from the QCRB in Eq. (\ref{eq:crb}). Consequently, the QFI $\mathcal{F}_T$ serves as the central figure of merit in most quantum thermometry studies. In practice, one aims to maximize the QFI as much as possible, thereby minimizing the achievable mean squared error $\mathrm{Var}[\mathcal{T}]$ according to the QCRB in Eq. (\ref{eq:crb}).

For the single-qubit probe considered here, the QFI can be evaluated using the practical Bloch‑vector representation~\cite{PhysRevA.87.022337,LiuJ.19.JPA} 
\begin{equation}\label{respF}
    \mathcal{F}_T(t)~=~|\partial_T \bm{r}|^2 + \frac{(\bm{r} \partial_T \bm{r})^2}{1-|\bm{r}|^2}.
\end{equation}
Here, $\bm{r}=(r_x,r_y,r_z)$ is the corresponding Bloch vector of the probe state $\rho_p$ satisfying the relation $\rho_p=(\mathrm{I}+\bm{r} \bm{\sigma})/2$, with $\mathrm{I}$ the $2\times 2$ identity matrix and $\bm{\sigma}=(\sigma_x,\sigma_y,\sigma_z)$ the vector of Pauli matrices. For the model given in Eq. (\ref{eq:master}), the Bloch‑vector components admit analytical expressions (see details in Appendix~\ref{a:4})
\bea\label{totalr}
    r_x(t) &=& \rho_{p,12}(0)\exp[\left(-2\gamma_0-\frac{1}{2}\gamma_p - i\omega\right)t] + \mathrm{H.c.}, \nonumber\\ 
    r_y(t) &=& i\rho_{p,12}(0)\exp[\left(-2\gamma_0-\frac{1}{2}\gamma_p - i\omega\right)t]+\mathrm{H.c.},\nonumber\\ 
    r_z(t) &=& r_z(0)e^{-\gamma_pt} + \frac{\gamma_m}{\gamma_p}(1-e^{-\gamma_pt}).
\eea
Here, $\mathrm{H.c.}$ denotes Hermitian conjugate, $\rho_{p,nm}(0)$ ($n,m=1,2$) are elements of an initial probe state $\rho_p(0)$ with $r_z(0)= \rho_{p,11}(0)-\rho_{p,22}(0)$. We have also introduced notions $\gamma_p\equiv \gamma_- + \gamma_+$. and $\gamma_m\equiv\gamma_+ - \gamma_-$. Substituting Eq.~\eqref{totalr} into Eq.~\eqref{respF}, we can get an analytical expression for the QFI $\mathcal{F}_T(t)$ of our probe model (see details in Appendix~\ref{a:4}),
\begin{widetext}
\begin{equation}
    \begin{aligned}
        \mathcal{F}_T(t) &= \frac{\gamma^2}{4T^4 \sinh^4\left(\frac{\omega}{2T}\right)}
        \left[t^2 e^{-(4\gamma_0+\gamma_p)t} |\rho_{p,21}(0)|^2 +\left(-r_z(0)te^{-\gamma_p t} -\frac{\gamma t e^{-\gamma_pt}}{\gamma_p} +\frac{\gamma (1 -e^{-\gamma_p t})}{\gamma_p^2} \right)^2 \right. \\ 
        &\left. + \frac{\left(2te^{-(4\gamma_0+\gamma_p)t}{|\rho_{p,21}(0)|^2}  
        -A(t)\right) ^2}
        {1-\left( 4 e^{-(4\gamma_0 + \gamma_p)t}|\rho_{p,21}(0)|^2 
        + \left|r_z(0)e^{-\gamma_p t} + \frac{\gamma_m}{\gamma_p} (1-e^{-\gamma_p t}) \right|^2 \right)} \right],
    \end{aligned}
    \label{analyF}
\end{equation}
with
\bea
A(t) &\equiv& \left[ \left(-r_z(0)^2+\frac{\gamma_m}{\gamma_p} r_z(0)\right) t  e^{-2\gamma_pt} -\frac{\gamma_m}{\gamma_p^2}r_z(0)e^{-\gamma_pt}(1-e^{-\gamma_pt}) \right. \nonumber\\ 
&&+ \left. \left(-r_z(0) \frac{\gamma_m}{\gamma_p} + \frac{\gamma_m}{\gamma_p^2}\right) te^{-\gamma_p t}(1-e^{-\gamma_pt}) -  \frac{\gamma_m^2}{\gamma_p^3}  (1-e^{-\gamma_pt})^2\right].  
\eea
\end{widetext}
From Eq. (\ref{analyF}), it is evident that pure dephasing affects the QFI $\mathcal{F}_T(t)$ only when the initial probe state carries nonzero coherences ($\rho_{p,21}(0)\neq 0$). Moreover, we note that terms of the QFI scale at most quadratically with time as expected. In the stationary limit of $t\to\infty$, the QFI approaches its stationary value
\begin{equation}\label{eq:F_T_s}
    \mathcal{F}_T^{\rm s}~=~\frac{\omega^2}{4T^4\cosh^2\left(\frac{\omega}{2T}\right)},
\end{equation}
which coincides exactly with the thermal QFI for the thermometer in a thermal equilibrium state determined by the sample (see details in Appendix \ref{a:5}).

In Fig.~\ref{fig:QFI}, we present a set of dynamical results for $\mathcal{F}_T(t)$ computed from Eq. (\ref{analyF}).
Several important observations emerge from these results: (i) Comparing the green (coherent initial state) and red (incoherent initial state) curves, we know that a coherent initial state yields a larger QFI $\mathcal{F}_T(t)$ over a substantial time interval. Since a larger QFI implies a lower achievable mean squared error and a higher precision, this reveals the beneficial role of quantum coherence in improving the precision of nonequilibrium thermometry, as observed in other thermometry settings~\cite{Ullah.23.PRR,Cruzeiro.24.E,Aiache.24.PRA,Jevtic.15.PRA}. Notably, the coherent case also exhibits $\mathcal{F}_T(t)>\mathcal{F}_T^{\rm s}$ for a wide range of times, demonstrating that a nonequilibrium thermometer can outperform its equilibrium counterpart with a smaller achievable mean squared error~\cite{Jevtic.15.PRA}. (ii) The green, orange, and blue curves illustrate how increasing the dephasing strength $\gamma_0$ reduces the magnitude of $\mathcal{F}_T(t)$ when the probe starts with coherence. In the limit of strong pure dephasing, $\mathcal{F}_T(t)$ approaches the value $\mathcal{F}^{\rm{in}}_T(t)$ obtained for an incoherent initial state. Because pure dephasing is ubiquitous in realistic settings, this trend implies that the aforementioned metrological advantage offered by quantum coherence is fragile and may vanish under practical noise conditions. In contrast, $\mathcal{F}^{\rm{in}}_T(t)$--derived from an incoherent initial state--remains insensitive to dephasing, as is evident from the analytical expression Eq.~(\ref{analyF}). Consequently, although an incoherent nonequilibrium thermometer lacks the enhancement provided by coherence, it offers robust performance against dephasing noise.

These findings confirm that the qubit-based probe adopted here performs on par with existing quantum thermometry models, validating its suitability as a nonequilibrium thermometer. Crucially, while this QFI analysis establishes the fundamental limits on mean squared error under realistic noise conditions, it assumes the existence of an unbiased estimator--a condition not automatically guaranteed in nonequilibrium settings. In the following, we move beyond the assessment of theoretical limits on the second-order mean squared error and demonstrate the practical utility of our direct temperature readout scheme, which provides a concrete strategy for obtaining direct temperature estimates from finite‑time nonequilibrium data.

%===============================================
\begin{figure}[b!]
 \centering
\includegraphics[width=1\columnwidth]{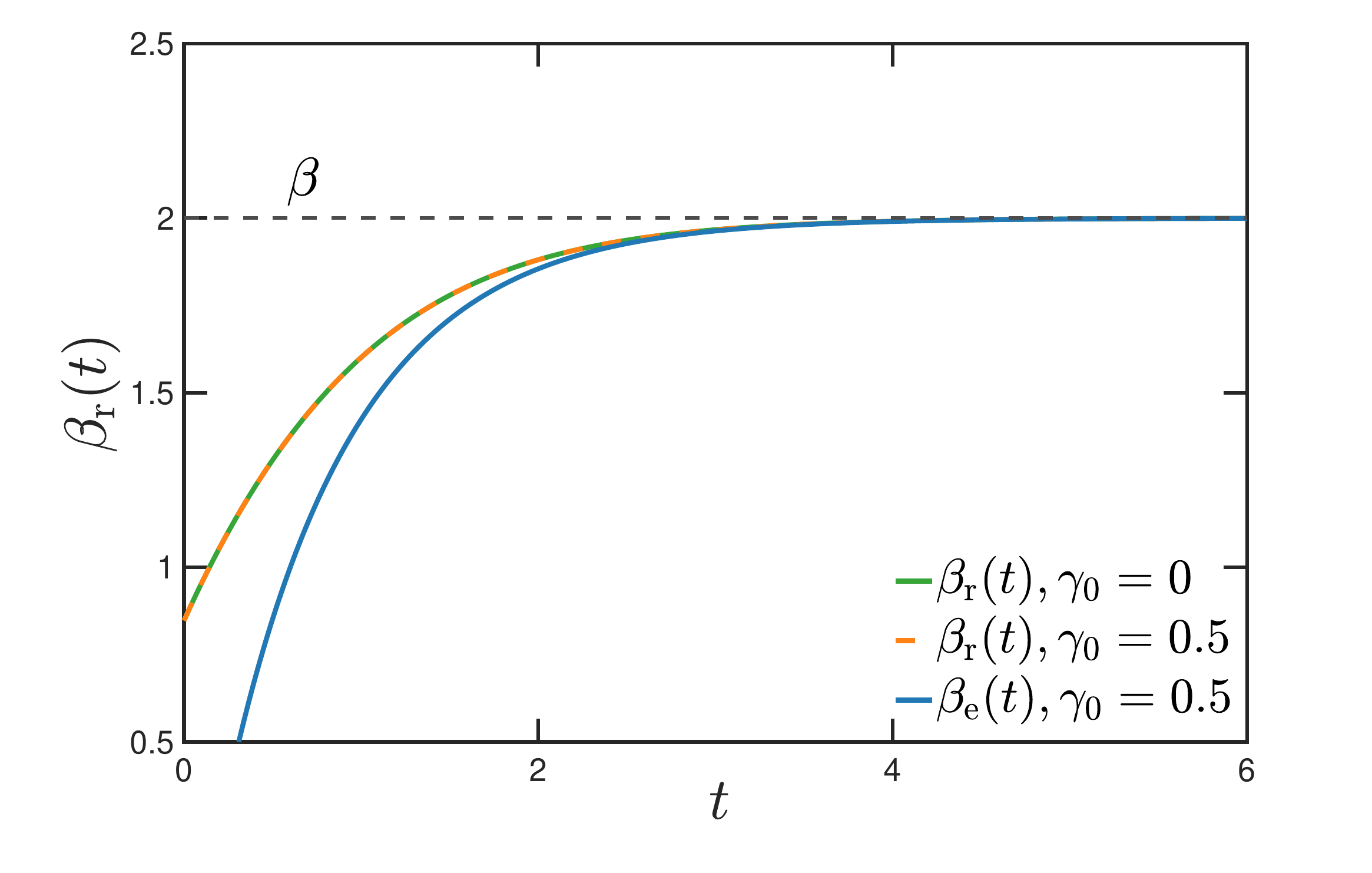} 
 \caption{Dynamics of the reference temperature $\beta_r(t)$ under a coherent initial state $\rho_0(0) = 0.5 \text{I} + 0.2 \sigma_x - 0.2 \sigma_z$ for different dephasing strengths $\gamma_0$. The blue solid line represents the effective temperature $\beta_e(t)$ for $\gamma_0=0.5$. The black dashed line marks the value of actual inverse temperature $\beta$. Other parameters are $\omega=1$, $T=0.5$ and $\gamma=1$.}
\protect\label{fig:direct}
\end{figure}
%===============================================
%==================================================================
\subsection{Behavior of reference temperature}\label{sec:reference}
%===============================================
\begin{figure*}[t!]
 \centering
\includegraphics[width=2\columnwidth]{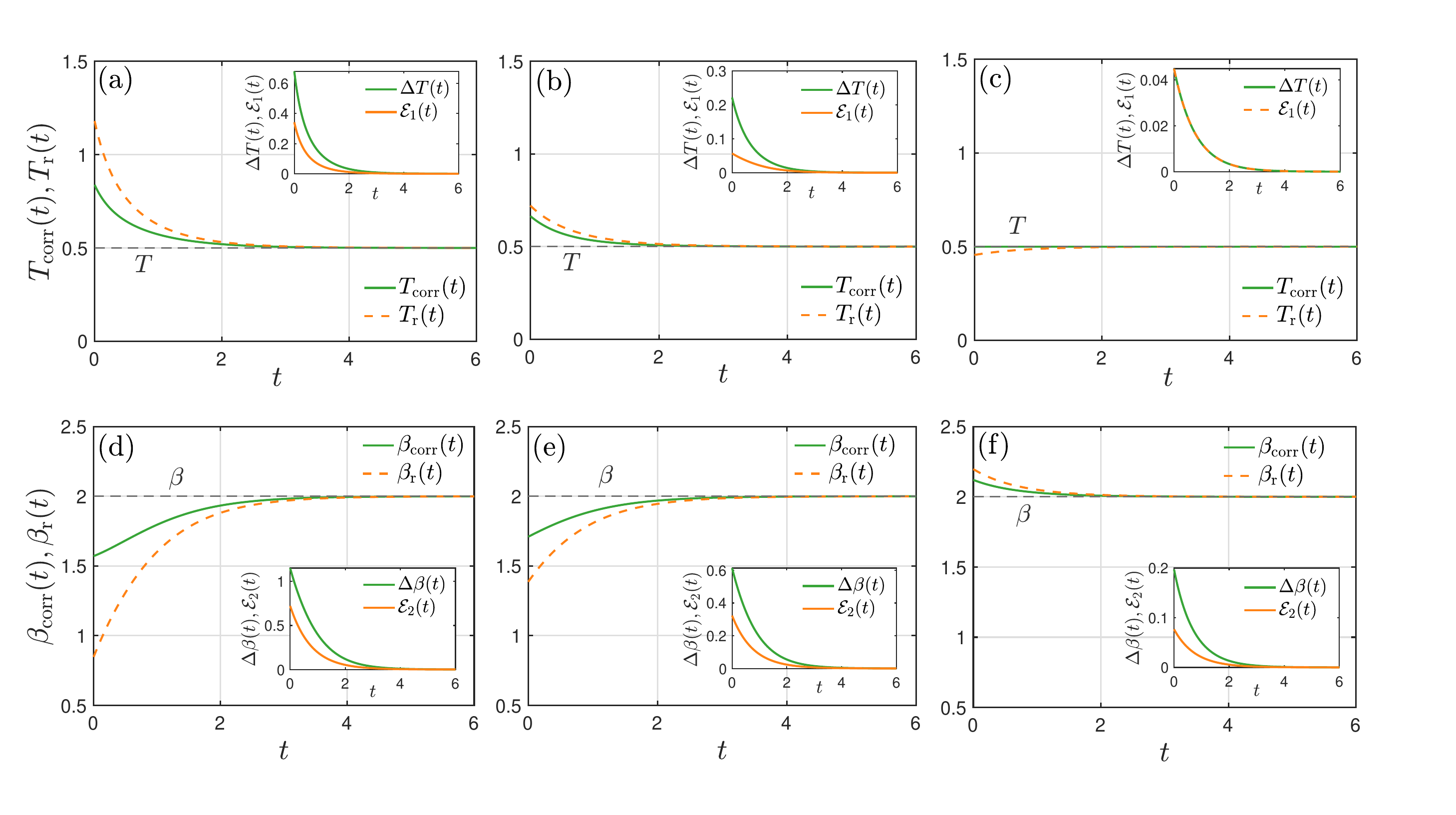} 
 \caption{Performance of the corrected dynamical temperatures $T_{\mathrm{corr}}(t)$ (upper panel, green solid line) and $\beta_{\mathrm{corr}}(t)$ (lower panel, green solid line) defined in Eqs. (\ref{eq:corr_T}) and (\ref{eq:corr_beta}), respectively.  All plots correspond to initial states with fixed coherence $\rho_{p,12}(0)=\rho_{p,21}(0)=0.2$ but varying populations: Left column (a,d) $\rho_{p,11}(0)=0.3,\rho_{p,22}(0)=0.7$; middle column (b,e) $\rho_{p,11}(0)=0.2,\rho_{p,22}(0)=0.8$; right column (c,f) $\rho_{p,11}(0)=0.1,\rho_{p,22}(0)=0.9$. For comparison, the reference temperature $T_{\mathrm{r}}(t)$ (upper panel, orange dashed line) and its inverse $\beta_{\mathrm{r}}(t)$ (lower panel, orange dashed line) are also shown. The horizontal black dashed line marks the actual temperature $T$ (upper panel) or its inverse $\beta$ (lower panel). Insets in the upper panel display the temperature deviation $\Delta T(t)=|T_r(t)-T|$ (green solid line) and its lower bound $\mathcal{E}_1(t)$ from Eq. (\ref{eq:epson1}) (orange solid line). Insets in the lower panel show the inverse-temperature deviation $\Delta\beta(t)=|\beta_r(t)-\beta|$ (green solid line) and its lower bound $\mathcal{E}_{2}(t)$ from Eq. (\ref{eq:epson2}) (orange solid line). Parameters are $\gamma_0=0$, $\omega=1$, $T=0.5$ and $\gamma=1$.}
\protect\label{fig:corre_T}
\end{figure*}
%===============================================
To access the performance of our direct temperature readout scheme, we begin by examining the behavior of the time-dependent reference temperature $\beta_r(t)$ which forms the foundation of our direct temperature readout scheme. To compute $\beta_r(t)$ numerically, we first evolve the master equation Eq.~\eqref{eq:master} to obtain the density matrix $\rho_p(t)$ of the probe and thereby its internal energy $E_p(t)$. Following the maximum entropy principle, we then introduce a Gibbsian reference state $\rho_r(t)$ parametrized by the reference temperature, and substitute it into Eq.~\eqref{eq:equal_e}. This procedure uniquely determines the value of $\beta_r(t)$ at each time step.

A set of dynamical results for $\beta_r(t)$ is illustrated in Fig.~\ref{fig:direct}. To maintain consistency with the preceding QFI analysis, we again consider a coherent initial state. The figure presents results for two dephasing strengths: $\gamma_0=0$ (green curve), representing the ideal dephasing-free case, and $\gamma_0=0.5$ (orange curve), corresponding to a realistic scenario with dephasing. For rigorous comparison, we also plot a conventionally-adopted effective temperature $\beta_e(t)$ (blue solid line), defined through the thermodynamic relation $\beta_e(t)=[\partial E_p(t)/\partial S(t)]^{-1}$, as a benchmark.

The results in Fig. \ref{fig:direct} reveal two important features. First, although $\beta_e(t)$ captures the overall trend of effective temperature evolution during thermal relaxation, the reference temperature $\beta_r(t)$ defined from the maximum entropy principle remains consistently closer to the true inverse sample temperature $\beta$ (marked by the black dashed line) at finite times. This confirms the superior estimation accuracy of $\beta_r(t)$ compared with $\beta_e(t)$ and its faster convergence to the true temperature over time, as anticipated by the inequality Eq. (\ref{eq:beta_dif}). Hence, the reference temperature carries clear thermodynamic significance as an effective temperature along nonequilibrium trajectories. Second, the magnitude of $\beta_{r}(t)$ remains unchanged regardless of the dephasing strength. This robustness stems from the fact that $\beta_r(t)$ is constructed solely from the energetic dynamics, which depend only on the populations of the thermometer's state in the energy basis and is therefore insensitive to dephasing. This behavior contrasts sharply with the QFI $\mathcal{F}_T(t)$, which is degraded by dephasing when initial coherence is present. The insensitivity to dephasing highlights the reliability and stability of the thermodynamic‑inference strategy in noisy environments.

\begin{figure*}[t!]
 \centering
\includegraphics[width=2\columnwidth]{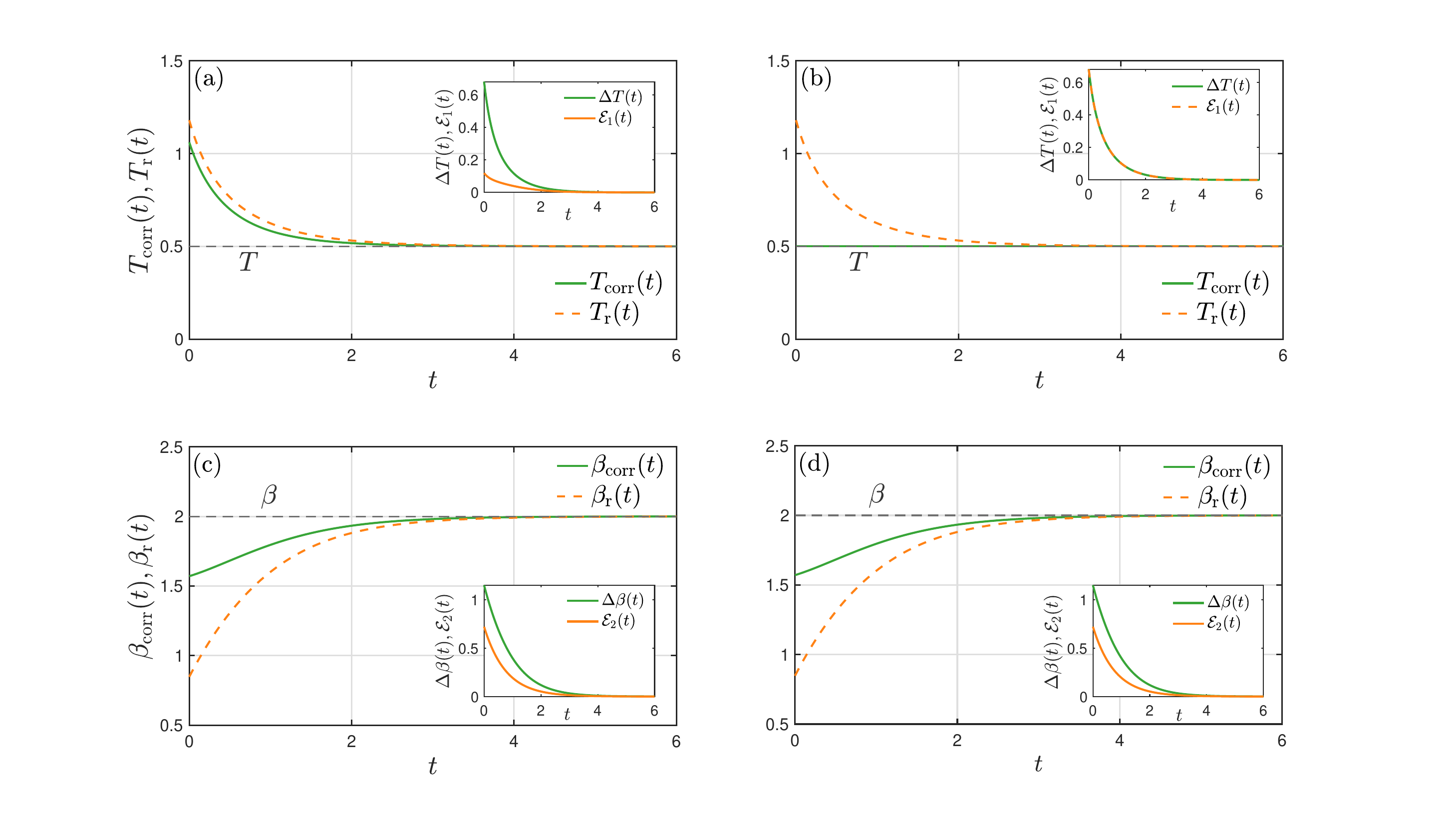} 
 \caption{ 
Performance of the corrected dynamical temperatures $T_{\mathrm{corr}}(t)$ (upper panel, green solid line) and $\beta_{\mathrm{corr}}(t)$ (lower panel, green solid line) defined in Eqs. (\ref{eq:corr_T}) and (\ref{eq:corr_beta}), respectively. All plots correspond to initial states with fixed populations $\rho_{p,11}(0)=0.3, \rho_{p,22}(0)=0.7$ but varying coherences: Left column (a, c) $\rho_{p,12}(0)=\rho_{p,21}(0)=0.3$; right column (b, d) $\rho_{p,12}(0)=\rho_{p,21}(0)=0.4$. For comparison, the reference temperature $T_{\mathrm{r}}(t)$ (upper panel, orange dashed line) and its inverse $\beta_{\mathrm{r}}(t)$ (lower panel, orange dashed line) are also shown. 
The horizontal black dashed line marks the actual temperature $T$ (upper panel) or its inverse $\beta$ (lower panel). Insets in the upper panel display the temperature deviation $\Delta T(t)=|T_r(t)-T|$ (green solid line) and its lower bound $\mathcal{E}_1(t)$ from Eq. (\ref{eq:epson1}) (orange solid line). 
Insets in the lower panel show the inverse-temperature deviation $\Delta\beta(t)=|\beta_r(t)-\beta|$ (green solid line) and its lower bound $\mathcal{E}_{2}(t)$ from Eq. (\ref{eq:epson2}) (orange solid line).Parameters are $\gamma_0=0$, $\omega=1$, $T=0.5$ and $\gamma=1$. 
}
\protect\label{fig:corre}
\end{figure*}
%===============================================
\subsection{Assessing temperature readout scheme}\label{sec:readout}
Having established the favorable properties of the reference temperature, we now examine the performance of the direct temperature readout scheme in a comprehensive manner. We will verify the lower bounds $\mathcal{E}_{1}(t)$ [cf. Eq. (\ref{eq:epson1})] and $\mathcal{E}_{2}(t)$ [cf. Eq. (\ref{eq:epson2})] on the temperature deviation $\Delta T(t)\equiv\left|T_{r}(t)-T\right|$ and the inverse-temperature deviation $\Delta\beta(t)\equiv\left|\beta_{r}(t)-\beta\right|$, respectively. We will also analyze the behavior of the corrected dynamical temperature $T_{\mathrm{corr}}(t)$ and its inverse counterpart $\beta_{\mathrm{corr}}(t)$ (Recalled that $\beta_{\mathrm{corr}}(t)\neq 1/T_{\mathrm{corr}}(t)$), which are designed to yield real‑time estimates that are more accurate and exhibit smaller bias than the raw reference temperatures $T_{r}(t)$ and $\beta_{r}(t)$. Particularly, we will systematically investigate how varying the initial state of the probe influences the precision of $T_{\mathrm{corr}}(t)$ and $\beta_{\mathrm{corr}}(t)$, thereby exploring initial‑state engineering as a means to further enhance the readout accuracy.%influences the accuracy of $T_{\mathrm{corr}}(t)$ and $\beta_{\mathrm{corr}}(t)$, thereby exploring initial‑state engineering as a means to further enhance the readout accuracy.

\subsubsection{Varying initial populations}
We first analyze the effect of varying the diagonal elements (populations) while keeping the off‑diagonal elements (coherences) fixed. A set of representative results is shown in Fig.~\ref{fig:corre_T}. Panels (a)-(c) show the dynamics of the corrected dynamical temperature $T_{\mathrm{corr}}(t)$ [cf. Eq.~\eqref{eq:corr_T}] for three initial states with increasing ground-state population. Comparing with the raw reference temperature $T_r(t)$, we see that $T_{\mathrm{corr}}(t)$ consistently yields a more accurate estimate of the actual temperature. Notably, when the initial populations are close to those of the true thermal state--as in Fig.~\ref{fig:corre_T}(c)--$T_{\mathrm{corr}}(t)$ provides an almost exact prediction despite the presence of nonzero initial coherence. The insets of panels (a)-(c) compare the temperature deviation $\Delta T = |T - T_{\mathrm{corr}}(t)|$ with the theoretical lower bound $\mathcal{E}_1(t)$ [cf. Eq.~\eqref{eq:epson1}], confirming that the actual error is strictly lower‑bounded by $\mathcal{E}_1(t)$ and eventually vanishes upon thermalization. 

Panels (d)-(f) display the corresponding results for the corrected dynamical inverse temperature $\beta_{\mathrm{corr}}(t)$ under the same initial states. The improvement of $\beta_{\mathrm{corr}}(t)$ over the raw reference inverse temperature $\beta_r(t)$ is also evident. The dependence on the initial population follows a trend similar to that of $T_{\mathrm{corr}}(t)$: estimates become more accurate as the initial population approaches equilibrium. However, the accuracy of $\beta_{\mathrm{corr}}(t)$ is lower than that of $T_{\mathrm{corr}}(t)$. Moreover, the insets of panels (d)-(f) verify the validity of the inequality Eq. (\ref{eq:beta_bound}).

Comparing the middle and right columns of Fig.~\ref{fig:corre_T} reveals an interesting feature: as the initial ground‑state population increases, the monotonicity of both $T_{\mathrm{corr}}(t)$ and $\beta_{\mathrm{corr}}(t)$ changes. Specifically, $T_{\mathrm{corr}}(t)$ shifts from a monotonic decrease to a monotonic increase, 
while $\beta_{\mathrm{corr}}(t)$ undergoes the opposite transition. Because the initial populations are varied with a relatively coarse spacing, this change in monotonicity indicates that the true temperature (or inverse temperature) lies within an interval bounded by two distinct $T_{\mathrm{corr}}(t)$ (or $\beta_{\mathrm{corr}}(t)$) trajectories. This observation underscores the practical utility of initial‑state engineering for refining temperature estimates.

\subsubsection{Varying initial coherence}
Following our analysis of diagonal-element variations, we now examine the specific role of initial quantum coherence in the performance of the direct temperature readout scheme. We expect that only $T_{\mathrm{corr}}(t)$ exhibits sensitivity to the magnitude of initial quantum coherence, which arises from the coherence dependence of the error bound $\mathcal{E}_1(t)$. In contrast, both the reference temperature $T_r(t)$ (or $\beta_r(t)$) and $\beta_{\rm{corr}}(t)$ are insensitive to the presence of initial quantum coherence. To isolate this effect, we fix the populations of the initial density matrix and systematically vary the magnitude of its off-diagonal elements (coherences). The corresponding numerical results are presented in Fig.~\ref{fig:corre}. Specifically, panels (a) and (c) correspond to an initial state with a coherence magnitude $|\rho_{p,12}(0)|=0.3$, while panels (b) and (d) correspond to a larger coherence magnitude $|\rho_{p,12}(0)|=0.4$.

As evidenced in Fig.~\ref{fig:corre} (a) and (b), increasing the initial quantum coherence consistently improves both the convergence rate and the final accuracy of the corrected temperature readout $T_{\mathrm{corr}}(t)$. This coherence-enhanced improvement in the temperature readout aligns with our earlier QFI analysis, where quantum coherence was shown to increase the magnitude of QFI of nonequilibrium thermometry at finite times. In contrast, the effect of coherence on the inverse-temperature readout $\beta_{\mathrm{corr}}(t)$ is markedly less pronounced. This distinct behavior stems from the distinct physical underpinnings of the two error functions. The temperature correction is governed by $\mathcal{E}_{1}(t)$, which is constructed from the quantum relative entropy $D(\rho_p || \rho_r)$ and the von Neumann entropy $S$; both quantities are directly sensitive to the coherence present in the probe's state. 
In contrast, the inverse-temperature correction $\mathcal{E}_{2}(t)$ depends only on the mean energy, $E_p(t)$, which--for the dynamics considered here--is insensitive to coherence at the level of expectation values. Consequently, while coherence substantially refines the temperature estimate, it offers little advantage for the inverse‑temperature readout based on $\beta_{\mathrm{corr}}(t)$. From the numerical results, we generally find that $T_{\rm{corr}}(t)$ achieves higher accuracy than $\beta_{\rm{corr}}(t)$ by harnessing initial quantum coherence.

\subsubsection{Evaluating iterative strategy}\label{sec:d3}
%===============================================
\begin{figure}[t!]
 \centering
\includegraphics[width=1\columnwidth]{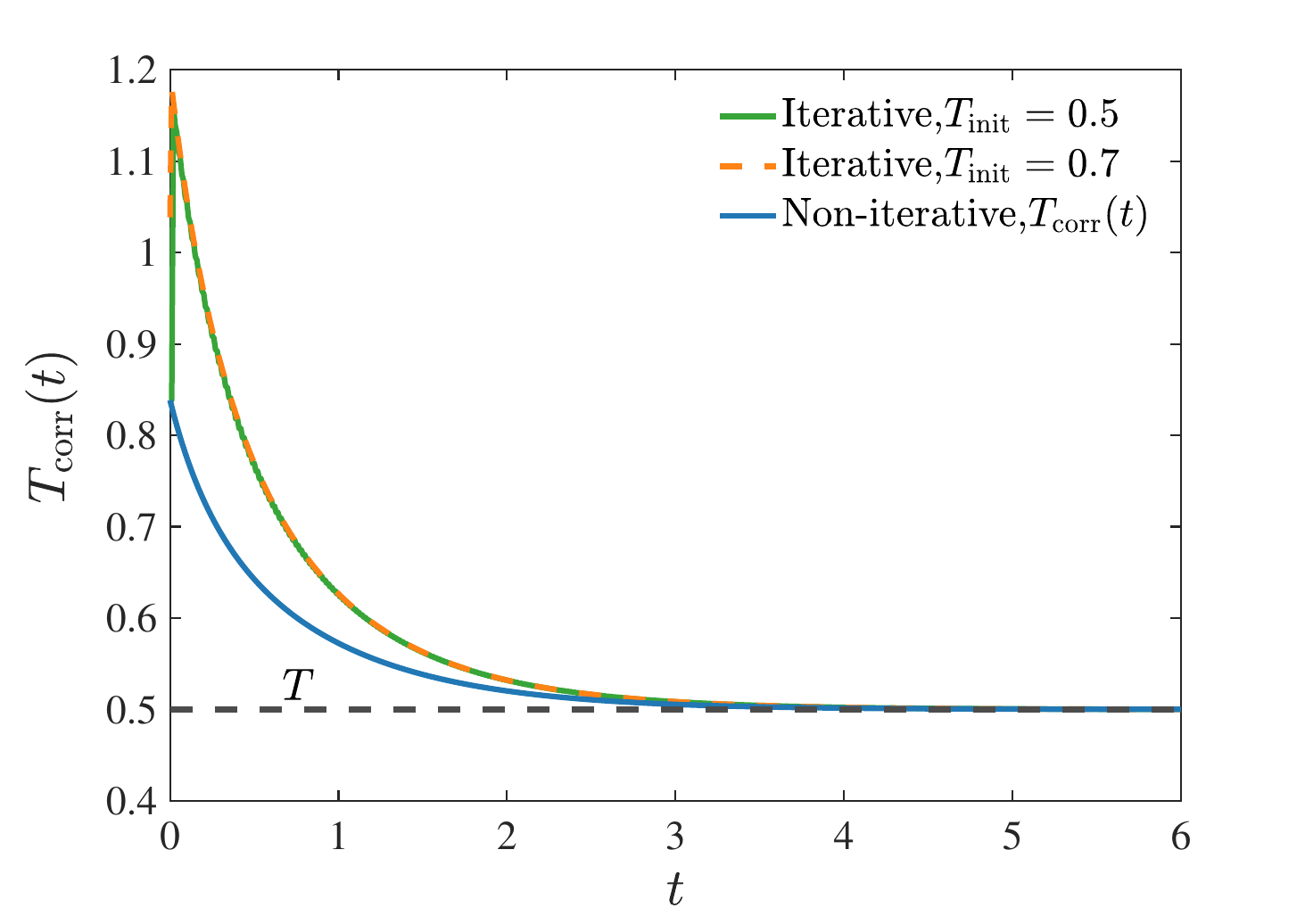} 
 \caption{Performance of the iterative scheme for calculating $T_{\mathrm{corr}}(t)$ with initial temperature guesses $T_{\rm{init}}=0.5$ (green solid line) and $T_{\rm{init}}=0.7$ (orange dashed line). For comparison, the blue solid line presents the result for $T_{\mathrm{corr}}(t)$ obtained without iteration. The black dashed line marks the value of actual temperature $T=0.5$. The initial state $\rho_p(0)$ and other parameters are the same as Fig. \ref{fig:corre_T} (a).}
\protect\label{fig:iteration}
\end{figure}
%===============================================
We now assume that prior knowledge of the actual temperature is unavailable and examine whether the iterative strategy proposed at the end of Sec.~\ref{sec:1} can still provide a reasonable temperature estimate. Fig.~\ref{fig:iteration} presents numerical results showing the evolution of $T_{\rm{corr}}(t)$ starting from two initial temperature guesses: $T_{\rm{init}}=0.5$ (green solid line) and $T_{\rm{init}}=0.7$ (orange dashed line). We note that the former guess is accurate, whereas the latter exhibits a notable deviation. Compared with the non-iterative result (blue solid line), which requires prior knowledge of the actual temperature, the iterative strategy generally yields a larger $T_{\rm{corr}}(t)$ at short times, even when the initial guess is accurate (green solid line). Notably, the iterative results appear to be independent of the initial temperature guess and consistently converge to the non-iterative result at later times, demonstrating the effectiveness and robustness of the iterative strategy.

%==================================================================
\section{Discussion and conclusion}\label{sec:3}
In this work, we have established a direct temperature readout framework that shifts the focus of nonequilibrium quantum thermometry from the analysis of quantum Fisher information to the provision of operationally accessible temperature estimates. By employing a thermodynamic inference strategy based on the maximum entropy principle, we first introduced a reference temperature $T_{r}(t)$ (or its inverse $\beta_r(t)$) and subsequently refined it using rigorously constructed error functions. These error functions serve as first‑order analogues of the quantum Cram\'er–Rao bound and quantify the estimation bias at finite times. Integrating these elements yields corrected dynamical temperatures $T_{\rm{corr}}(t)$ and $\beta_{\rm{corr}}(t)$ that provide the final postprocessed temperature readouts. This construction guarantees the convergence of the temperature readout to the true temperature upon thermalization at weak couplings. Using a qubit‑based thermometer as a specific example, we clarified the physical resources and practical utility of the scheme and demonstrated that initial state engineering can improve the accuracy of the direct temperature readout. 

We note that evaluating these error functions formally requires knowledge of the true temperature, a requirement inherent to the local thermometry setting. In practice, however, the sample temperature is often confined to an approximate interval, such as that set by the operating range of a cryostat. Within such an interval, the error functions can be computed iteratively. This approach preserves the scheme's experimental viability without requiring exact prior knowledge of the temperature. As a result, our framework offers a versatile tool for real‑time thermal monitoring in emerging quantum technologies, ranging from quantum computing platforms to nanoscale thermal management.

\section*{Acknowledgments}
J.L. acknowledges support from the National Natural Science Foundation of China (Grant No. 12205179), the Shanghai Pujiang Program (Grant No. 22PJ1403900) and the Shanghai Science and Technology Innovation Action Plan (Grant No. 24LZ1400800).

\appendix
%=============================================================
\renewcommand{\theequation}{A\arabic{equation}}
\renewcommand{\thefigure}{A\arabic{figure}}
\setcounter{equation}{0}  % reset counter
\setcounter{figure}{0}  % reset counter
\section{Comparing $\beta_r(t)$ with $\beta_e(t)$}
\label{a:1}
 In this appendix, we prove inequality Eq.~(\ref{eq:beta_dif}) in the main text, which states that the reference inverse temperature $\beta_r(t)$ inferred from the maximum-entropy principle yields a more accurate estimate of the true inverse temperature $\beta$ than the commonly used effective temperature $\beta_e(t) \equiv [\partial E_p(t)/\partial S(t)]^{-1}$--a direct generalization of the equilibrium definition to nonequilibrium settings. Here, $E_p(t)$ and $S(t)$ denote the instantaneous internal energy and von Neumann entropy of the probe, respectively. 
 
 Our derivation rests on the asymptotic behavior of Markovian thermal relaxation processes, as described, for instance, by the quantum Lindblad master equation Eq.~(\ref{eq:master}) in the main text. In such processes, the time‑evolving probe state satisfies
\begin{equation}\label{eq:limit}
    \lim_{t\to\infty}\rho_p(t) = \rho_{T}.
\end{equation}
Here, $\rho_p(t)$ represents the time-evolving probe state, and $\rho_T = e^{-\beta H_p}/Z_T$ with $Z_T=\mathrm{Tr}[e^{-\beta H_p}]$ denotes the stationary thermal state of the probe at inverse temperature $\beta$ of the thermal sample.

As the system temporally evolves towards equilibrium $\rho_T$, the Gibbsian reference state $\rho_r(t)$--strictly determined by the instantaneous probe energy--monotonically approaches this thermal state. This convergence implies that the quantum relative entropy between the reference state and the final thermal state at different times should satisfy the following inequality 
\begin{equation}\label{eq:d_in}
    D[\rho_r(t+\tau)||\rho_T] \le D[\rho_r(t)||\rho_T],
\end{equation}
where $\tau \ge 0$ represents a non-negative time lag. Given that both $\rho_r(t)$ and $\rho_T$ are diagonal in the energy eigenbasis of $H_p$, the relative entropy explicitly simplifies to:
\begin{equation}\label{eq:dd}
   D[\rho_r(t)||\rho_T] = \ln\frac{Z_T}{Z_t} - [\beta_r(t)-\beta]E_p(t). 
\end{equation}
In getting the above equation, we have utilized the property satisfied by the Gibbsian reference state, $E_p(t)=\mathrm{Tr}[H_p\rho_p(t)]=\mathrm{Tr}[H_p\rho_r(t)]$.

Inserting Eq. (\ref{eq:dd}) into Eq. (\ref{eq:d_in}) and arranging terms, we arrive at
\bea\label{eq:b4}
\beta [E_p(t)-E_p(t+\tau)] &\ge& \ln \frac{Z_r(t)}{Z_r(t+\tau)} + \beta_r(t)E_p(t) \notag \\
&& - \beta_r(t+\tau)E_p(t+\tau).
\eea
We now introduce
\bea\label{eq:est_1}
\beta_1(t, \tau) &\equiv& \frac{1}{E_p(t+\tau) - E_p(t)} \Bigg\{ \ln \left[ \frac{Z_r(t+\tau)}{Z_r(t)} \right] \notag \\
&&+ \beta_r(t+\tau)E_p(t+\tau) - \beta_r(t)E_p(t) \Bigg\}.
\eea
In the limit of $\tau\to 0$, we have $\beta_1(t,\tau\to0)=\beta_r(t)$. With $\beta_1(t,\tau)$, Eq. (\ref{eq:b4}) implies the following relative relations,
\begin{equation}\label{eq:re_1}
    \left\{\begin{array}{ll}
    \beta~\ge~\beta_{1}(t,\tau),~~\mathrm{When}~E_p(t)-E_p(t+\tau)> 0,\\
    \beta~\le~\beta_{1}(t,\tau),~~\mathrm{When}~E_p(t)-E_p(t+\tau)< 0.
    \end{array}\right.
\end{equation}

For the probe at finite times, we can define its generalized nonequilibrium free energy $\mathcal{F}(t)$~\cite{Liu.23.PRAa} as
\begin{equation}\label{eq:nfe}
    \mathcal{F}(t)~=~E_p(t)-T_r(t)S(t).
\end{equation}
After a straightforward derivation, we can find that $\mathcal{F}(t)=-T_r(t)\ln Z_r(t)+T_r(t)D[\rho_p(t)||\rho_r(t)]$ which, combining with Eq. (\ref{eq:nfe}), yields
\begin{equation}\label{eq:b8}
    \beta_r(t)E_p(t)+\ln Z_r(t)~=~D[\rho_p(t)||\rho_r(t)]+S(t).
\end{equation}
Inserting Eq. (\ref{eq:b8}) into Eq. (\ref{eq:est_1}), we get 
\begin{equation}\label{eq:connection}
    \begin{aligned}
        \beta_{1}(t,\tau) &= \frac{S(t+\tau)-S(t)}{E_p(t+\tau)-E_p(t)}+\frac{\Delta D[\rho_p || \rho_r]}{E_p(t+\tau)-E_p(t) }  \\
    &= \beta_{2}(t,\tau)+\frac{\Delta D[\rho_p || \rho_r]}{E_p(t+\tau)-E_p(t)}.
    \end{aligned}
\end{equation}
where
\begin{equation}
    \begin{aligned}
        \Delta D[\rho_p || \rho_r]  &= D[\rho_p(t+\tau)||\rho_r(t+\tau)]-D[\rho_p(t)||\rho_r(t)].  \\  
    \end{aligned}
\end{equation}
In the last line, we have denoted
\begin{equation}
    \beta_{2}(t,\tau)~\equiv~\left(\frac{\Delta E_p(t,\tau)}{\Delta S(t,\tau)}\right)^{-1}.
\end{equation}
Here, $\Delta E_p(t,\tau)=E_p(t+\tau)-E_p(t)$ and $\Delta S(t,\tau)=S(t+\tau)-S(t)$. In the limit of $\tau\to 0$, we find $\beta_2(t,\tau\to0)=\beta_e(t)$.

Since $\rho_p(t+\tau)$ is closer to a Gibbsian form than $\rho_p(t)$ in Markovian thermal relaxation processes described by Eq. (\ref{eq:master}), we expect $D[\rho_p(t)||\rho_r(t)]\ge D[\rho_p(t+\tau)||\rho_r(t+\tau)]$ in Eq. (\ref{eq:connection}). Therefore, we infer that $\beta_{1}(t,\tau)\ge \beta_{2}(t,\tau)$ ($\beta_{1}(t,\tau)\le \beta_{2}(t,\tau)$) when $E_p(t)-E_p(t+\tau)>0$ ($E_p(t)-E_p(t+\tau)<0$). Combining with Eq. (\ref{eq:re_1}), we get
\begin{equation}\label{eq:re_2}
    \left\{\begin{array}{cc}
    \beta~\ge~\beta_{1}(t,\tau)~\ge~\beta_{2}(t,\tau),~~\mathrm{When}~E_p(t)-E_p(t+\tau)>0,\\
    \beta~\le~\beta_{1}(t,\tau)~\le~\beta_{2}(t,\tau),~~\mathrm{When}~E_p(t)-E_p(t+\tau)<0.
    \end{array}\right.
\end{equation}
Taking the limit of $\tau \to 0$, we have $\beta_1(t,\tau)\to\beta_r(t)$ and $\beta_2(t,\tau)\to\beta_e(t)$, 
we then conclude that $\beta_{r}(t)$ is always more accurate than $\beta_{e}(t)$ in estimating the actual inverse temperature $\beta$ as stated by Eq. (\ref{eq:beta_dif}) in the main text.

%=============================================================
\renewcommand{\theequation}{B\arabic{equation}}
\renewcommand{\thefigure}{B\arabic{figure}}
\setcounter{equation}{0}  % reset counter
\setcounter{figure}{0}  % reset counter
\section{Lower bound on temperature deviation $|T_r(t)-T|$}
\label{a:2}
In this appendix, we prove inequality Eq. (\ref{eq3}) of the temperature deviation $|\Delta T| = |T_{r}(t) - T|$ in the main text. We consider a probe with Hamiltonian $H_p$ in a nonequilibrium state $\rho_p(t)$. The main concept we utilize is the generalized nonequilibrium free energy in Eq. (\ref{eq:nfe}). By denoting $F_r(t)=-T_r(t)\ln Z_r(t)=E_p(t)-T_r(t)S_r(t)$ ($S_r(t) = -\Tr(\rho_r(t)\ln\rho_r(t))$) which is the free energy associated with the reference Gibbsian state, we can rewrite Eq. (\ref{eq:nfe}) as 
\begin{comment}
At time $t$, the energy and von Neumann entropy of the probe are given by
\begin{equation}
E_p(t) \equiv \Tr(H\rho_p(t)),\quad
S_p(t) \equiv -\Tr(\rho_p(t)\ln\rho_p(t)).
\end{equation}
Following the maximum-entropy principle, we introduce a Gibbs reference state
\begin{equation}
\rho_r(t) = \exp(-\frac{H_p}{T_r(t)})/{Z_r(t)},
\end{equation}
where the normalization coefficient $Z_r = \Tr(e^{-H_p/T_r(t)})$ and the reference temperature $T_r$ at each time is determined by the principle of equal energy,
\begin{equation}
\Tr(H_p\rho_r(t)) = \Tr(H_p\rho_p(t)) .
\end{equation}
The von Neumann entropy of the reference state is denoted by
$S_r(t) = -\Tr(\rho_r(t)\ln\rho_r(t))$.
A standard identity in quantum thermodynamics relates the free-energy difference to the quantum relative entropy,
\end{comment}
\begin{equation}
\mathcal{F}(t) - F_r(t) = T_r(t) D[\rho_p(t)\|\rho_r(t)],
\label{D4}
\end{equation}
For later convenience, we further introduce the Helmholtz free energy associated with the final thermal equilibrium state $\rho_T$
\begin{equation}
    F_{T} = E_{T} - T S_{T}.
\end{equation}
Here, $E_{T} \equiv \Tr(H_p\rho_T)$ and $S_{T}=-\mathrm{Tr}[\rho_T\ln\rho_T]$ is the internal energy of the probe at thermal equilibrium and the von Neumann entropy of $\rho_T$, respectively. 

We first have
\begin{equation}
F_{T} - F_{r}(t) ~=~ T_{r}(t) S_{r}(t) - T S_{T}+E_{T}-E_p(t).
\end{equation}
Since $\mathcal{F}(t)- F_{r}(t) = \left[ \mathcal{F}(t) - F_{T} \right] + \left[ F_{T} - F_{r}(t) \right]$, we find
\bea
T_{r}(t) S_{r}(t) - T S_{T} &=& T_{r}(t) D(\rho_p(t)\|\rho_{r}(t)) \notag \\
&& - (\mathcal{F}(t) - F_{T}) + E_p(t) - E_{T}\nonumber\\
&=& T_{r}(t) D(\rho_p(t)\|\rho_{r}(t))+T_r(t)S(t)\nonumber\\
&&-TS_{T}.
\eea
From the last line of the above equation, we get
\bea
    [T_r(t) - T] S_r(t) &=& T_r(t) D[\rho_p(t)\| \rho_r(t)] \notag 
    \\
    &&+ [T_r(t) S(t) - T S_r(t)],
    \label{D9}
\eea
or equivalently, 
\bea\label{eq:b6}
    T_r(t) - T &=& \frac{T_r(t) D[\rho_p(t)\| \rho_r(t)]}{S_r(t)} \notag 
    \\
    &&+ \left[T_r(t) \frac{S(t)}{S_r(t)} - T \right].
    \label{D9}
\eea
Taking the absolute value of both sides of Eq.~\eqref{eq:b6} and applying the triangle inequality $|A+B| \geq \big||A|-|B|\big|$, we obtain
\begin{equation}
|T_r(t) - T| \geq \left| \left|\frac{T_r(t) D[\rho_p(t) \| \rho_r(t)]}{S_r(t)}\right| - \left|T_r\frac{S(t)}{S_r(t)}-T\right|\right|.
\end{equation}
Noting that the term $T_r(t) D[\rho_p(t) \| \rho_r(t)]/S_r(t)$ is always positive, we thus recover Eq. (\ref{eq3}) in the main text.

\begin{comment}
Therefore, we define the error function $\varepsilon_1$ as:
\begin{equation}
\epsilon_1(t) \equiv \Big|\,\frac{T_r(t) D[\rho_p(t) \| \rho_r(t)]}{S_r(t)} - \frac{|T_r(t) S(t) - T S_r(t)|}{S_r(t)} \Big|\,\geq 0 .
\end{equation}
Finally, we obtain a lower bound for the reference temperature $T_r$:
\begin{equation}
|T_r(t) - T| \geq \mathcal{E}_1(t) .
\label{D12}
\end{equation}
The bound in Eq.~\eqref{D12} is obtained without any assumption of proximity between $T_r$ and $T$, and therefore applies to general nonequilibrium situations.
Moreover, the error function $\mathcal{E}_1$ vanishes in the thermalized limit $\rho \to \rho_r$, where $D(\rho\|\rho_r)\to 0$ and $S\to S_r$.
Finally, $\epsilon_1$ should be interpreted as a thermodynamic lower bound on the temperature deviation, rather than as a metric distance between quantum states.
\end{comment}

%=============================================================
\renewcommand{\theequation}{C\arabic{equation}}
\renewcommand{\thefigure}{C\arabic{figure}}
\setcounter{equation}{0}  % reset counter
\setcounter{figure}{0}  % reset counter
\section{Lower bound on inverse-temperature contrast $|\beta_r(t)-\beta|$}
\label{a:3}
In this appendix, we prove inequality Eq. (\ref{eq:beta_bound}) of the inverse-temperature deviation $|\beta_r(t)-\beta|$ in the main text. To proceed, we define an interpolating inverse temperature linear in $\beta_r(t)-\beta$,
\begin{equation}
    \beta_s~\equiv~\beta+s[\beta_r(t)-\beta].
\end{equation}
Here, $s\in[0,1]$ such that $\beta_{0}=\beta$ and $\beta_1=\beta_r(t)$. We further assign a Gibbsian state with respect to $\beta_s$
\begin{equation}
    \rho_g^s~\equiv~\frac{e^{-\beta_sH_p}}{Z_s}
\end{equation}
with $Z_s=\mathrm{Tr}[e^{-\beta_sH_p}]$. We have $\rho_g^0=\rho_T$ and $\rho_g^1=\rho_r(t)$.

As we consider inferring temperature from energy measurements, we invoke the following relation
\bea\label{eq:contrast_def}
E_T-E_p(t) &=& -\mathrm{Tr}\left[\int_0^1\frac{d}{ds}\left(H_p\rho_g^s\right)ds\right].
\eea
Here, we have denoted $E_T=\mathrm{Tr}[H_p \rho_g^0]$ and utilized the relation $E_p(t)=\mathrm{Tr}[H_p\rho_g^1]=\mathrm{Tr}[H_p\rho_r(t)]$ according to Eq. (\ref{eq:equal_e}) in the main text. Since the derivative on the right-hand-side of Eq. (\ref{eq:contrast_def}) can be expanded as
\begin{align}
\frac{d}{ds}\left(H_p\rho_g^s\right) &= [\beta_r(t)-\beta] H_p \Big( \rho_g^s \mathrm{Tr}[H_p\rho_g^s]  - H_p\rho_g^s \Big).
\end{align}
%\bea
%\frac{d}{ds}\left(H_p\rho_g^s\right) &=& H_p\Big(-[\beta_r(t)-\beta]H_p\rho_g^s+\rho_g^s[\beta_r(t)-\beta]\mathrm{Tr}[H_p\rho_g^s]\Big).
%\eea
We can rewrite Eq. (\ref{eq:contrast_def}) as
\bea\label{eq:contrast}
E_T-E_p(t) &=& [\beta_r(t)-\beta]\int_0^1\,\mathrm{Cov}_{\rho_g^s}(H_p,H_p)ds.
\eea
Here, we have defined a covariance 
\begin{equation}
    \mathrm{Cov}_{\rho}(A,B)=\mathrm{Tr}[AB\rho]-\mathrm{Tr}[A\rho]\mathrm{Tr}[B\rho].
\end{equation}
We rearrange Eq. (\ref{eq:contrast}) to get the following expression
\begin{equation}\label{eq:beta_contrast}
    \beta_r(t)-\beta~=~\frac{E_T-E_p(t)}{\int_0^1\,\mathrm{Cov}_{\rho_g^s}(H_p,H_p)ds}.
\end{equation}

We can bound the inverse-temperature contrast $\beta_r(t)-\beta$ from below by noting that one can use Schatten-$p$ norms for operators to bound from above the covariance. For a given operator $A$, the corresponding Schatten-$p$ norms are defined as
\begin{equation}
    ||A||_p~\equiv~\left(\sum_l(\alpha_l)^{p}\right)^{\frac{1}{p}}.
\end{equation}
Here, $p\in[1,\infty)$ and singular values $\{\alpha_l\}$ are the eigenvalues of $\sqrt{A^{\dagger}A}$. For later convenience, we denote the operator norm $||A||_{\infty}=\max_l|a_l|$ and the trace norm $||A||_1=\sum_la_l$. Denoting $\bar{A}=A-\mathrm{Tr}[\rho A]$, we have
\bea\label{eq:bound_cov}
   |\mathrm{Cov}_{\rho}(A,B)| &=& |\mathrm{Tr}[\rho \bar{A}B]|~\le~||\rho \bar{A}B||_1\nonumber\\
   &\le& ||\rho||_1||\bar{A}B||_{\infty}=||\bar{A}B||_{\infty}\nonumber\\
   &\le& ||\bar{A}||_{\infty}||B||_{\infty}=||A||_{\infty}||B||_{\infty}.
\eea
In getting the first line, we have used an inequality for the trace norm $|\mathrm{Tr}[A]|~\le~||A||_1$. In arriving at the second line, we have used the H\"older's inequality $||AB||_p~\le~||A||_{q_1}||B||_{q_2}$ with $1/p=1/q_1+1/q_2$ by setting $p=1,q_1=1,q_2=\infty$ and the fact that $||\rho||_1=1$. To get the third line, we have used the H\"older's inequality with $p=\infty,q_1=q_2=\infty$ and $||\bar{A}||_{\infty}=||A||_{\infty}$. Inserting Eq. (\ref{eq:bound_cov}) into Eq. (\ref{eq:beta_contrast}), we finally get
\bea\label{eq:bound_1}
    |\beta_r(t)-\beta| &\ge& \frac{|E_T-E_p(t)|}{(||H_p||_{\infty})^2}.
\eea 
This is just Eq. (\ref{eq:beta_bound}) in the main text.

%=============================================================
\renewcommand{\theequation}{D\arabic{equation}}
\renewcommand{\thefigure}{D\arabic{figure}}
\setcounter{equation}{0}  % reset counter
\setcounter{figure}{0}  % reset counter
\section{Evaluating quantum Fisher information for qubit-based thermometer}\label{a:4}
In this appendix, we present derivation details that lead to the analytical expression Eq. (\ref{analyF}) of quantum Fisher information (QFI) showed in the main text. In the Bloch sphere representation, a qubit state can be written as
\begin{equation} \label{relationshiprho}
    \rho_p(t) = \frac{1}{2}\left[\mathbb{I} + \bm{r}(t)\cdot \bm{\sigma}\right].
\end{equation}
where $\bm{r} = (r_x,r_y,r_z)^T$ is the Bloch vector and $\bm{\sigma} = (\sigma_x,\sigma_y,\sigma_z)$ denotes the Pauli matrices. Since $\partial_tr_i(t) = \text{Tr}[\partial_t\rho_p(t) \sigma_i]$, we can obtain the following equations of motion for elements of the Bloch vector based on the quantum master equation~(\ref{eq:master}) in the main text (time dependence is suppressed),
\begin{equation}
    \begin{aligned}
        \partial_tr_x&=-\frac{1}{2}(4\gamma_0+\gamma_p)r_x-\omega r_y, \\
     \partial_tr_y &=\omega r_x-\frac{1}{2}(4\gamma_0+\gamma_p)r_y, \\
     \partial_tr_z & =\gamma_m - r_z\gamma_p.
    \end{aligned}
\end{equation}
Here, we have introduced notations $\gamma_p = \gamma_-+ \gamma_+$ , $\gamma_m = \gamma_- - \gamma_+$. By solving this set of equations of motion under a given initial condition $\rho_p(0)$, we can obtain an analytical solution for $\bm{r}$,
\begin{equation}\label{eq:a3}
    \begin{aligned}
        r_x(t) &= \rho_{p,12}(0)\exp[\left(-2\gamma_0-\frac{1}{2}\gamma_p - i\omega\right)t] + \mathrm{H.c.}, \\ 
        r_y(t) &= i\rho_{p,12}(0)\exp[\left(-2\gamma_0-\frac{1}{2}\gamma_p - i\omega\right)t]+\mathrm{H.c.},\\ 
        r_z(t) &= r_z(0)e^{-\gamma_pt} + \frac{\gamma_m}{\gamma_p}(1-e^{-\gamma_pt}),
    \end{aligned}
\end{equation}
which is just Eq.~\eqref{totalr} in the main text. To obtain an analytical solution for QFI based on Eq. (\ref{eq:a3}), we need to express $\partial_T \bm{r}$ and $\bm{r}\partial_T \bm{r}$,
\begin{equation}\label{eq:a4}
    \begin{aligned}
        \partial_T r_x(t) &= -\frac{1}{2}\gamma_p't e^{-(2\gamma_0+\frac{1}{2}\gamma_p)t}[\rho_{21}(0)e^{i\omega t} +\mathrm{H.c.}] \\ 
        \partial_T r_y(t) &= \frac{1}{2}\gamma_p't e^{-(2\gamma_0+\frac{1}{2}\gamma_p)t}[i\rho_{21}(0)e^{i\omega t} +\mathrm{H.c.}] \\ 
        \partial_T r_z(t) &= -\gamma_p't e^{-\gamma_pt}\left(r_z(0) - \frac{\gamma_m}{\gamma_p}\right) 
        + (1 - e^{-\gamma_pt})\left( \frac{\gamma_m}{\gamma_p} \right)'.
    \end{aligned}
\end{equation}
Here, we have denoted $\gamma_{m,p}'\equiv \partial_T\gamma_{m,p}$ the derivative of damping coefficients $\gamma_{m,p}$ with respect to the temperature $T$. Particularly, we have $\gamma_p' =2\gamma \omega e^{\frac{\omega}{T}}/[T^2 (e^{\frac{\omega}{T}}-1)^2]$ and $\gamma_m'=0$ according to their expressions. We have also implicitly assumed that the initial state is independent of $T$ which should be the case in general. With Eqs. (\ref{eq:a3}) and (\ref{eq:a4}), by utilizing the relation $\left( \frac{\gamma_m}{\gamma_p} \right)' = \frac{\gamma_m'}{\gamma_p} - \frac{\gamma_m}{\gamma_p^2}\gamma_p'$, we obtain the following expression after a straightforward derivation,
\begin{equation}\label{eq:a5}
\begin{aligned}
    \bm{r}\cdot\partial_T \bm{r} 
        &= \gamma_p' \Bigg[ -2te^{-(4\gamma_0+\gamma_p)t}|\rho_{21}(0)|^2 -  \frac{\gamma_m^2}{\gamma_p^3}  (1-e^{-\gamma_pt})^2  \\
         &+  \left(-r_z(0)^2+\frac{\gamma_m}{\gamma_p} r_z(0)\right) t  e^{-2\gamma_pt} \\
         &-\frac{\gamma_m}{\gamma_p^2}r_z(0)e^{-\gamma_pt}(1-e^{-\gamma_pt})\\ 
        & + \left(-r_z(0) \frac{\gamma_m}{\gamma_p} + \frac{\gamma_m}{\gamma_p^2}\right) te^{-\gamma_pt}(1-e^{-\gamma_pt}) \Bigg].
\end{aligned}
\end{equation}

Inserting expressions Eqs. (\ref{eq:a3})-(\ref{eq:a5}) into the definition of QFI Eq.~\eqref{respF}, we can get the analytical expression for QFI shown in Eq.~\eqref{analyF} of the main text.

%=============================================================
\renewcommand{\theequation}{E\arabic{equation}}
\renewcommand{\thefigure}{E\arabic{figure}}
\setcounter{equation}{0}  % reset counter
\setcounter{figure}{0}  % reset counter
\section{Thermal quantum Fisher information}\label{a:5}
When the probe reaches thermal equilibrium with the sample, its QFI, dubbed thermal QFI, reads (see, e.g., Refs.~\cite{Hovhannisyan.21.PRXQ,LiuJ.19.JPA})
\begin{equation}
    \mathcal{F}_{\rm{th}}~=~\frac{C}{T^2}.
\end{equation}
Here, $T$ denotes the sample's temperature and $C\equiv d(\mathrm{Tr}[\rho_TH_p])/dT$ is the heat capacity of the probe. For the probe's Hamiltonian $H_p=\omega \sigma_z/2$, we can readily calculate
\begin{equation}
    \mathrm{Tr}[\rho_TH_p]~=~-\frac{\omega}{2}\tanh\left(\frac{\omega}{2T}\right).
\end{equation}
which yields $C=\frac{\omega^2}{4T^2\cosh^2\left(\frac{\omega}{2T}\right)}$. Hence, we find
\begin{equation}
    \mathcal{F}_{\rm{th}}~=~\frac{\omega^2}{4T^4\cosh^2\left(\frac{\omega}{2T}\right)}.
\end{equation}

%\bibliography{thermo}

\end{document}